# Integrated microring resonator and waveguide polarizers based on partially photo-reduced 2D graphene oxide thin films

*David J. Moss* [1,*]

[1] Optical Sciences Centre, Swinburne University of Technology, Melbourne, 3122, Australia

*dmoss@swin.edu.au*

**Abstract**

Optical polarizers, which selectively transmit light with specific polarization states, are essential components in modern optical systems. Here, we experimentally demonstrate integrated waveguide and microring resonator (MRR) polarizers incorporating reduced graphene oxide (rGO). 2D graphene oxide (GO) films are integrated onto silicon photonic devices with precise control over their thicknesses and sizes, followed by GO reduction via two different methods including uniform thermal reduction and localized photothermal reduction. We measure devices with different lengths, thicknesses, and reduction degrees of the GO films. The results show that the devices with rGO exhibit better performance than those with GO, achieving a polarization-dependent loss of ~47 dB and a polarization extinction ratio of ~16 dB for the hybrid waveguides and MRRs with rGO, respectively. By fitting the experimental results with theory, it is found that rGO exhibits more significant anisotropy in loss, with an anisotropy ratio over 4 times that of GO. In addition, rGO shows higher thermal stability and greater robustness to photothermal reduction than GO. These results highlight the strong potential of rGO films for implementing high-performance polarization selective devices in integrated photonic platforms.

**Keywords**: integrated optics, 2D materials, graphene oxide, optical polarizers

**Introduction**

In modern optical systems, controlling light polarization is of fundamental importance and underpins a variety of advanced optical technologies [1-3]. Optical polarizers, which allow the transmission of light with a specific polarization orientation and block light with the orthogonal polarization, are key components for controlling light polarization [4]. To date, a wide range of optical polarizers have been implemented based on refractive prisms [5, 6], birefringent crystals [7, 8], fiber components [9, 10], and integrated photonic devices [11-13]. However, these polarizers based on bulk materials typically face challenges in achieving effective polarization selection across broad wavelength ranges [1, 14, 15], despite the growing demand for broadband optical polarizers driven by rapid advancements in photonic technologies and systems [16, 17].

Since the groundbreaking isolation of graphene in 2004 [18], there has been an enormous surge in research on two-dimensional (2D) materials with atomic-scale thicknesses, which exhibit many extraordinary properties unattainable for conventional bulk materials [19-21]. With highly anisotropic properties across wide optical bands, 2D materials such as graphene [15, 22, 23], graphene oxide (GO) [24-26], and transition metal dichalcogenides (TMDCs) [27-29] have been incorporated into bulk material device platforms to realize high-performance optical polarizers. As a common derivative of graphene, GO has facile solution-based synthesis processes and transfer-free film coating with precise control over the film thickness, making it well-suited for large-scale on-chip integration to implement hybrid devices [30-32]. In addition, the properties of GO can be easily changed through various reduction methods, providing a high flexibility to optimize the performance of hybrid devices [33-35].

Previously, we demonstrated integrated optical polarizers incorporating 2D GO films on both doped silica and silicon device platforms [24, 25]. In this work, we integrate 2D reduced GO (rGO) films onto integrated photonic devices to realize waveguide and MRR polarizers with improved performance. We fabricate hybrid integrated devices with precise control over the thicknesses and lengths of the GO films. The reduction of GO is realized by using two methods: uniform thermal reduction, achieved by heating the integrated chip on a hot plate, and localized photothermal reduction, induced by high power of input light. Detailed measurements are carried out for devices with different lengths, thicknesses, and reduction degrees of the GO films. The results show that the devices with rGO exhibit better polarization selectivity than comparable devices with GO. Up to ~47-dB polarization-dependent loss (*PDL*) and ~16-dB polarization extinction ratio (*PER*) are achieved for the hybrid waveguides and MRRs with rGO, respectively. By fitting the experimental results with theoretical simulations, we find that rGO exhibits significantly improved loss anisotropy, with an anisotropy ratio more than 4 times that of GO. Compared to GO, rGO also exhibits stronger thermal stability and lower sensitivity to photothermal reduction. These results verify the effectiveness of on-chip integration of 2D rGO films to realize high performance optical polarizers.

## Results and discussion

### *Device design and fabrication*

As an oxidized derivative of graphene, GO consists of carbon networks decorated with various oxygen functional groups (OFGs), such as hydroxyl, epoxide, carbonyl, and carboxylic groups [33, 36, 37]. **Figure 1(a)** illustrates the atomic structures and bandgaps of graphene oxide (GO), semi-reduced GO (srGO), and highly reduced GO (hrGO). Due to the presence of isolated $sp^2$ domains within the $sp^3$ carbon-oxygen matrix, unreduced GO is a dielectric material with an opened bandgap of ~2.1 − 3.6 eV [36, 38]. This bandgap is larger than the energy of two photons at ~1550 nm (*i.e.*, ~1.6 eV), allowing for both low linear light absorption and two-photon absorption at infrared wavelengths. The reduction of GO breaks the chemical bonds between the OFGs and the carbon network. Compared to pristine GO, reduced GO (rGO) has a decreased bandgap [39, 40], resulting in alterations to material properties such as refractive index, optical absorption, and electrical conductivity. Practically, the reduction of GO films can be achieved by using different methods, such as thermal reduction, chemical reduction, and photoreduction [41-43]. As the degree of reduction increases, the fraction of $sp^2$-hybridized carbon atoms increases. For hrGO with minimal remaining OFGs, the bandgap and material properties closely resemble those of graphene, which exhibits a zero bandgap and metallic behaviour [44, 45].

**Figure 1(b)** shows the schematic of an integrated waveguide polarizer based on a silicon photonic waveguide coated with a 2D GO film. The cross section of the silicon waveguide is 400 nm × 220 nm. The GO film has a thickness of 4 nm, which corresponds to 2 layers of GO fabricated using a solution-based self-assembly method (as discussed later in this section). **Figure 1(c)** shows the corresponding transverse electric (TE) and transverse magnetic (TM) mode profiles for the hybrid waveguide in **Fig. 1(b)**, which were simulated using a commercial mode-solving software (COMSOL Multiphysics). The TE- and TM polarized effective indices (at 1550 nm) for the hybrid waveguide were $2.093 + 1.244 \times 10^{-4}i$ and $2.093 + 4.784 \times 10^{-5}i$, respectively. In our simulation, the refractive index ($n$) and extinction coefficient ($k$) of GO for TE polarization were $n_{TE} = \sim 1.969$ and $k_{TE} = \sim 0.0098$, respectively. For TM polarization, the corresponding values were $n_{TM} = \sim 1.898$ and $k_{TM} = \sim 0.0022$. The $n$, $k$ values were obtained from our previous measurements in Ref. [34] and the experimental results in following sections. The large difference between $k_{TE}$ and $k_{TM}$ is due to the significant anisotropy in the light absorption of 2D GO films, where the in-plane absorption is much stronger than the out-of-plane absorption [15, 24]. As a result, TE-polarized (in-plane) light experiences a higher loss compared to TM-polarized (out-of-plane) light as it propagates through the hybrid waveguide, allowing the hybrid waveguide to function as a TM-pass optical polarizer.

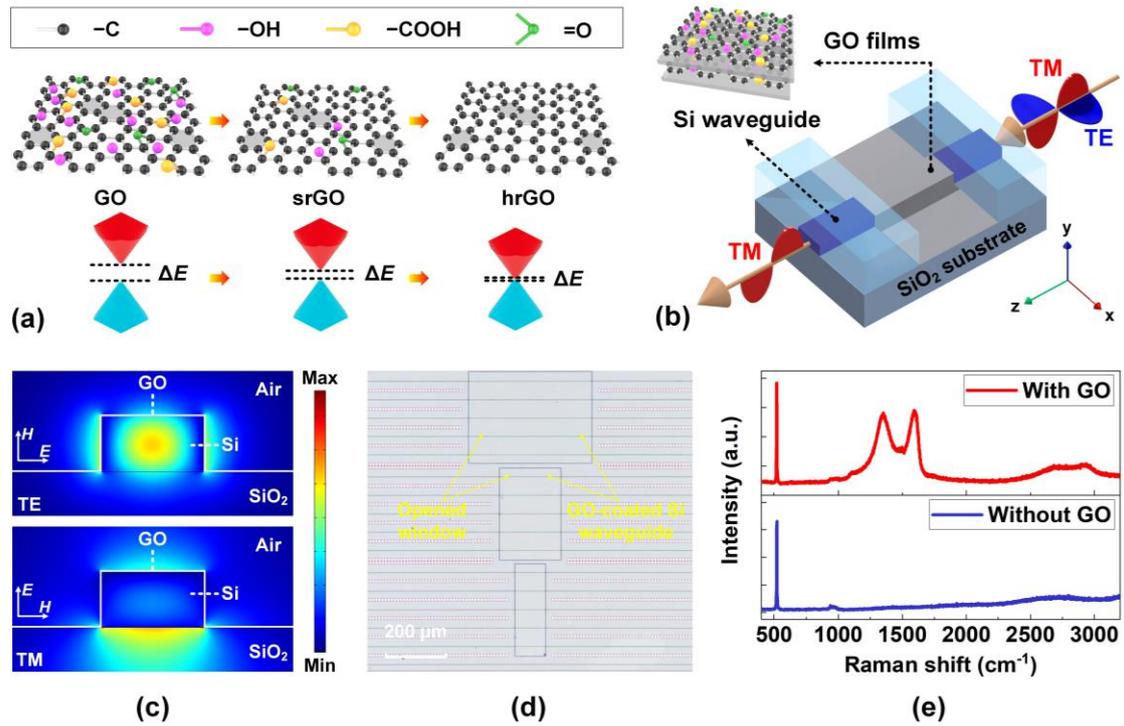

**Fig. 1** | (**a**) Schematics of atomic structures and bandgaps of graphene oxide (GO), semi-reduced GO (srGO), and highly reduced GO (hrGO). (**b**) Schematic illustration of a GO-coated silicon waveguide as an optical polarizer. Inset illustrates the layered GO film structure fabricated by self-assembly. (**c**) TE and TM mode profiles for the hybrid waveguide with 2 layers of GO. (**d**) Microscopic image of the fabricated devices on a GO-coated silicon-on-insulator (SOI) chip. (**e**) Measured Raman spectra of the SOI chip in (**d**) without GO and with 2 layers of GO.

**Figure 1(d)** shows a microscopic image of the fabricated devices on a silicon-on-insulator (SOI) chip. The silicon waveguides were patterned via 248-nm deep ultraviolet lithography followed by inductively coupled plasma etching. After this, a 1.5-μm-thick silica layer was deposited by plasma enhanced chemical vapor deposition to cover the SOI chip as an upper cladding. To enable the interaction between the GO films and the evanescent field from the silicon waveguides, windows were opened on the silica upper cladding to allow the coating of 2D GO films onto the silicon waveguides. In our fabricated devices, the length of all silicon waveguides was ~3.0 mm, and the lengths of the opened windows ranged between ~0.1 mm and ~2.2 mm.

The coating of the GO film, with a thickness of ~2 nm per layer, was realized by using a solution-based self-assembly method that enabled transfer-free and layer-by-layer film coating [24, 30]. During the coating process, a multilayered film structure composed of alternating GO layers and oppositely charged polymer layers was constructed, with the GO layers formed through the self-assembly of exfoliated GO nanoflakes. As compared with the complicated film transfer methods employed for other 2D materials such as graphene and transition-metal dichalcogenides (TMDCs) [46, 47], this coating method allows for transfer-free film coating with precise control of the film thickness. In addition, it enables conformal coating of 2D GO films onto silicon waveguides with minimal air gaps [48]. In **Fig. 1(d)**, the coated GO film shows high transmittance and good morphology without any noticeable wrinkling or stretching, confirming excellent film attachment onto the silicon waveguides.

**Figure 1(e)** shows the measured Raman spectra of the SOI chip in **Fig. 1(d)** before and after coating 2 layers of GO, which were measured using a ~514-nm pump laser. The GO films had a thickness of ~4.0 nm, which was characterized using atomic force microscopy measurement. In the Raman spectrum for the GO-coated chip, the existence of the representative $D$ (~1345 cm$^{-1}$) and $G$ (~1590 cm$^{-1}$) peaks [49, 50] provides evidence for successful on-chip integration of 2D GO film.

*Polarization-dependent loss measurements*

In **Fig. 2**, we show the measured insertion losses ($IL$'s) of our fabricated devices for input continuous-wave (CW) light in different polarization states. We measured devices with different GO film lengths ($L_{GO}$) and GO layer numbers ($N$), after being subjected to various reduction temperatures ($T_R$) ranging from ~50 to ~200 °C. Here we chose the temperature range of $T_R \leq 200$ °C because the polymer layers in the self-assembled films cannot withstand temperatures beyond this range. For all the devices, the cross section of the uncoated silicon waveguides was ~400 nm × 220 nm. In our measurements, lensed fibers were employed to butt couple a CW light at ~1550 nm into and out of the fabricated devices with inverse-taper couplers at both ends. The fiber-to-chip coupling loss was ~5 dB per facet. For comparison, we measured the *IL*'s by using the same input power of ~0 dBm. Unless otherwise specified, the input power ($P_{in}$) and *IL* in our following discussion refers to those measured after excluding the fiber-to-chip coupling loss.

**Figures 2(a-i)** and **2(a-ii)** show the measured TE- and TM-polarized *IL* versus $L_{GO}$ for the hybrid waveguides with 1 layer of GO ($N$ = 1), respectively. Before the *IL* measurement, the SOI chip was heated on a hot plate for 15 minutes at various temperatures $T_R$. For comparison, the results corresponding to different $T_R$ are plotted together with those measured at room temperature prior to heating (which are labeled as 'initial'). In each figure, the data points represent the average values from measurements of three duplicate devices, and the error bars reflect the variations across different samples. We do not show results for *IL* > 70 dB in these and subsequent figures because it exceeds the detection range of the optical power meter used in our measurements.

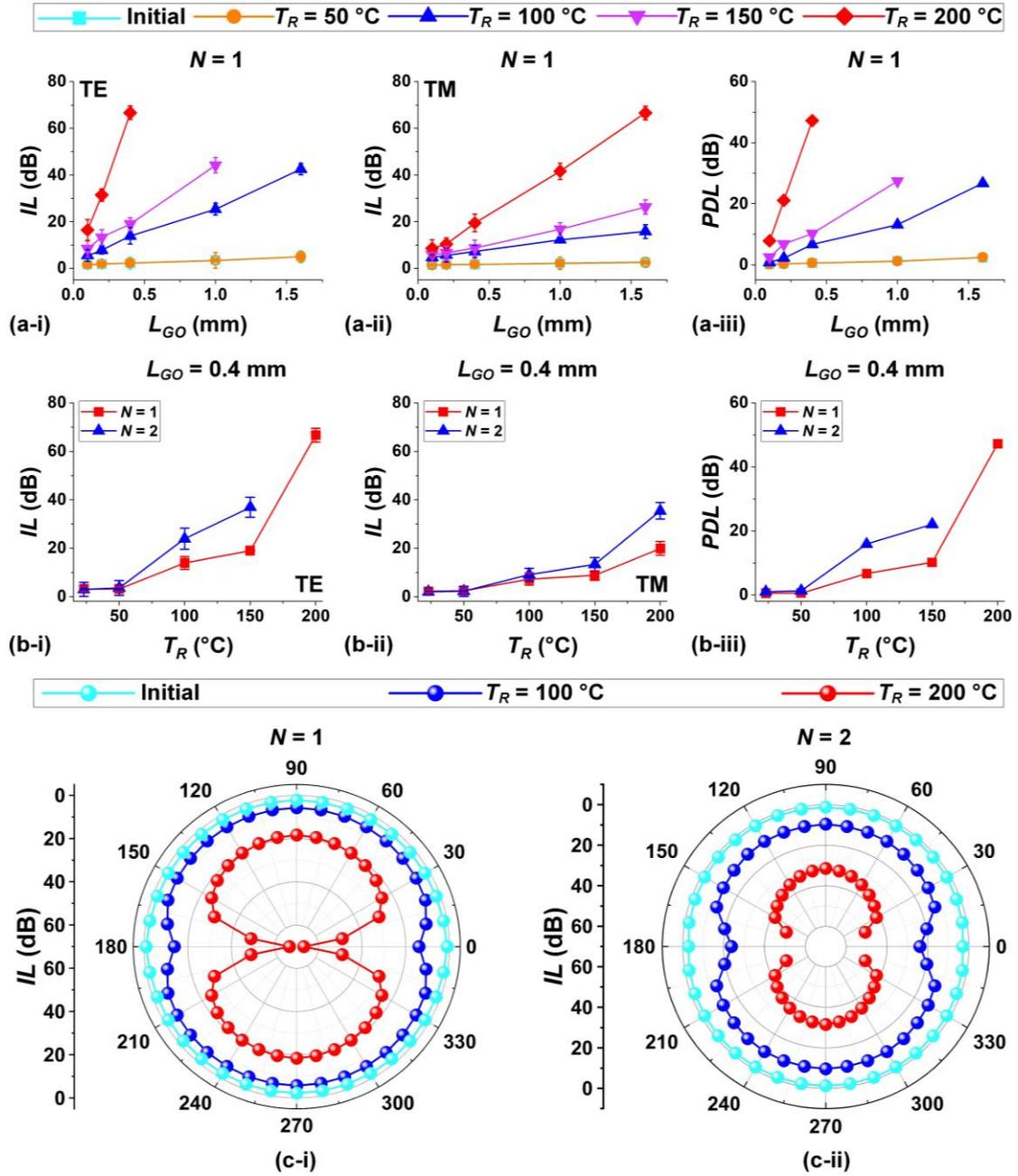

**Fig. 2** | (**a**) Measured (**i**) TE- and (**ii**) TM-polarized insertion loss ($IL$) versus GO coating length ($L_{GO}$) for the hybrid waveguides coated with a monolayer GO film ($N$ =1) after the chip was heated at various temperatures $T_R$. (**iii**) shows the polarization dependent loss ($PDL$) calculated from (**i**) and (**ii**). (**b**) Measured (**i**) TE- and (**ii**) TM- polarized $IL$ versus $T_R$ for the hybrid waveguides with 1−2 layers of GO ($N$ =1, 2). (**iii**) shows the $PDL$ calculated from (**i**) and (**ii**). (**c**) Polar diagrams for the measured $IL$ of devices with different GO layer numbers of (**i**) $N$ = 1 and (**ii**) $N$ = 2 after the chip was heated at various temperatures $T_R$. The polar angle represents the angle between the input polarization plane and the substrate. In (**a**) – (**c**), the input continuous-wave (CW) power and wavelength were ~0 dBm and ~1550 nm, respectively. In (**a**) and (**b**), the data points illustrate the average of measurements on three duplicate devices and the error bars depict the variations among the different devices. In (**b**) and (**c**), $L_{GO}$ = ~0.4 mm.

In **Fig. 2(a-i)** and **2(a-ii)**, the *IL* increases with $L_{GO}$ for both polarizations, with the TE polarization exhibiting a more dramatic increase than the TM polarization. At $T_R$ = ~50 °C, both the TE- and TM-polarized *IL* did not show any significant difference as compared with that at the initial unheated status. These results suggest that there were no significant changes in the GO film properties at $T_R$ = ~50 °C, indicating that the reduction of GO did not occur at this temperature. In contrast, when $T_R \geq$ ~100 °C, the *IL* increases with $T_R$ for both polarizations. This reflects the loss increase due to the reduction of GO at high temperatures. As $T_R$ increases, a higher degree of reduction was achieved, leading to a more significant increase in the *IL*.

**Figure 2(a-iii)** shows the corresponding *PDL* (dB) obtained by subtracting the TM-polarized *IL* in **Fig. 2(a-ii)** from the TE-polarized *IL* in **Fig. 2(a-i)**. The *PDL* increases with $L_{GO}$, and it also increases with $T_R$ when $T_R \geq$ 100 °C. For the device with $L_{GO}$ = ~0.4 mm and at $T_R$ = ~200 °C, a maximum *PDL* value of ~47 dB was obtained. In contrast, the *PDL* exhibited no significant difference between the initial status and at $T_R$ = 50 °C, achieving a *PDL* of ~1 dB for the devices with the same $L_{GO}$. By further increasing $L_{GO}$ for devices with hrGO (*i.e.*, at $T_R$ = ~150 and ~200 °C), a *PDL* exceeding ~47 dB can be achieved (not shown in this figure due to limited detection range of the optical power meter), at the expense of a higher additional *IL* induced by GO.

**Figures 2(b-i)** and **2(b-ii)** show the measured *IL* versus $T_R$ for TE and TM polarizations, respectively. Here we show the results for the hybrid waveguides with 1 and 2 layers of GO. For comparison, all the waveguides had the same $L_{GO}$ = ~0.4 mm. Both the TE- and TM-polarized *IL* remains unchanged when $T_R \leq$ ~50 °C. For $T_R \geq$ ~100 °C, the *IL* for TE polarization shows a more significant increase with $T_R$ than that for TM polarization, following a trend similar to that in **Fig. 2(a-i)** and **2(a-ii)**.

Compared to the devices with 1 layer of GO ($N$ = 1), higher *IL* was achieved for the devices with 2 layers of GO ($N$ = 2), reflecting a higher loss induced by a thicker GO film. **Figure 2(b-iii)** shows the corresponding *PDL* extracted from **Fig. 2(b-i)** and **2(b-ii)**, where higher *PDL* values were also achieved for the devices with thicker GO films. For the 2-layer device at $T_R$ = ~150 °C, the *PDL* was ~ 24 dB, in contrast to ~10 dB for a comparable 1-layer device. At $T_R$ = ~200 °C, it is anticipated that the 2-layer device can achieve a high *PDL* exceeding 60 dB, we were not able to measure it due to the limited detection range of our optical power meter.

**Figures 2(c-i)** and **2(c-ii)** show the polar diagrams for the measured *IL* of devices with 1 and 2 layers of GO ($N$ = 1, 2), respectively. In each figure, we plot three curves corresponding to different $T_R$. For comparison, all the hybrid waveguides had the same $L_{GO}$ = ~0.4 mm. The polar diagrams show variations in *IL* values across different polarization angles, which reflects the polarization selectivity of the hybrid waveguides. For the hybrid waveguides with 1 layer of GO, the *PDL* values at the initial unheated status, $T_R$ = ~100 °C, and $T_R$ = ~200 °C are ~1 dB, ~7 dB, and ~47 dB, respectively. These results indicate that the polarization selectivity is improved as the degree of GO reduction increases. At $T_R$ = ~100 °C, the *PDL* values for $N$ = 1 and $N$ = 2 are ~7 dB and ~15 dB, respectively. This reflects that improved polarization selectivity can also be achieved for hybrid devices with thicker GO films.

*Analysis of GO film properties*

Based on the measured results in **Fig. 2**, we further analyze the properties of 2D GO films by fitting the experimental results with theoretical simulations. **Figure 3(a)** shows the waveguide propagation loss (*PL*) versus $T_R$ for the hybrid devices with 1 and 2 layers of GO (*i.e.*, $N$ = 1, 2), which was extracted from the measured *IL* in **Fig. 2(b-i)** and **2(b-ii)**. The excess propagation loss (*EPL*) induced by the GO films was further calculated by excluding the *PL* for the uncoated silicon waveguide (*i.e.*, ~3.4 dB/cm for TE polarization and ~3.1 dB/cm for TM polarization). The TE-polarized *EPL* induced by 1 layer of rGO at $T_R$ = 200°C was ~1520 dB/cm, in contrast to ~20 dB/cm for 1 layer of unreduced GO at the initial status. This reflects the substantial increase in loss for highly reduced GO films. We also note the value of ~1520 dB/cm is lower than the typical values of *EPL* induced by monolayer graphene (*i.e.*, ~2000 dB/cm [51, 52]). This suggests that, although the GO film was highly reduced at $T_R$ = 200°C, it was not yet fully reduced.

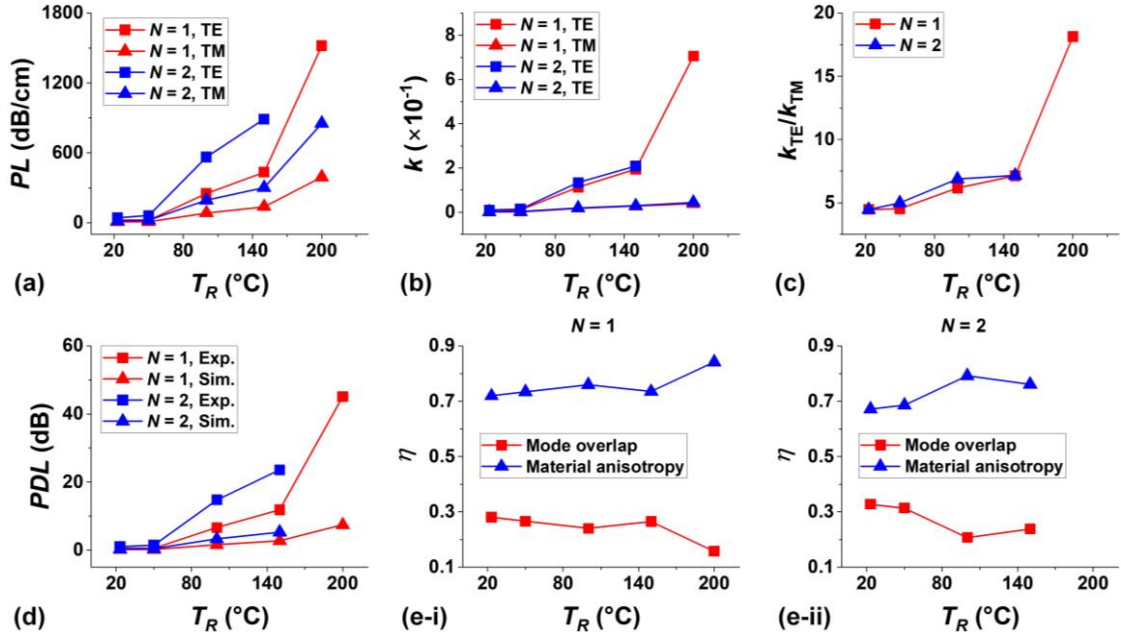

**Fig. 3** | (**a**) TE- and TM-polarized waveguide propagation loss (*PL*) versus $T_R$ for the hybrid waveguides with 1 and 2 layers of GO (*N* = 1, 2). (**b**) Extinction coefficients (*k*'s) of 2D GO films versus $T_R$ obtained by fitting the results in (**a**) with optical mode simulations. (**c**) Anisotropy ratios of *k* values for TE and TM polarizations ($k_{TE} / k_{TM}$) extracted from (**b**). (**d**) Measured (Exp.) and simulated (Sim.) *PDL* versus $T_R$ for the hybrid waveguides with 1-2 layers of GO (*N* = 1, 2). The simulated *PDL* values were obtained by using the same *k* value for both TE and TM polarizations. (**e**) Fractional contributions (*η*'s) to the overall *PDL* from polarization-dependent mode overlap and material loss anisotropy, which were extracted from (**d**). (**i**) and (**ii**) show the results for *N* = 1 and 2, respectively.

**Figure 3(b)** shows the extinction coefficients (*k*'s) of 2D GO films obtained by fitting the results in **Fig. 3(a)** with optical mode simulations (at 1550 nm) for the hybrid waveguides. For 1 layer of rGO at $T_R$ = ~200 °C, the value of *k* is ~0.7057 for TE polarization, which is about 75 times that of comparable unreduced GO. For all different $N$ and $T_R$, the GO films exhibited higher values of $k$ for TE polarization than TM polarization, reflecting the intrinsic anisotropy in the loss of 2D GO films. We also note that, for both polarizations, slightly higher $k$ values were obtained with thicker GO films. This is likely due to the increased scattering loss caused by film unevenness and accumulation of imperfect contact between adjacent layers in thicker films.

In **Fig. 3(c)**, we further plot the anisotropy ratios defined as the ratios of the corresponding $k$ values for TE- and TM- polarizations ($k_{TE} / k_{TM}$) in **Fig. 3(b)**. Compared to unreduced GO, higher values of the anisotropy ratio are obtained for rGO at $T_R \geq$ 100 °C, with the anisotropy ratio increasing as the degree of reduction increases. For 1 layer of rGO at $T_R = $ ~200 °C, the anisotropy ratio is ~18 – over 4 times higher than that of 1 layer of unreduced GO. These results highlight an interesting phenomenon that the 2D GO films exhibit more significant loss anisotropy as the degree of reduction increases. This is probably because the reduction of GO leads to the removal of OFGs and hence a decrease in the film thickness. We also note that in Ref. [15] monolayer graphene (with a thickness of ~0.5 nm, in contrast to ~2 nm for monolayer unreduced GO) exhibits a higher anisotropy ratio of ~30. This suggests that highly reduced GO exhibits loss anisotropy close to that of graphene.

Compared to GO, the higher anisotropy ratio of rGO leads to more significant difference between the absorption of in-plane and out-of-plane light waves, making it better suited for implementing optical polarizers with high polarization selectivity. In addition, unlike the intricate film transfer methods needed for on-chip integration of graphene, GO offers advantages for large-scale manufacturing due to its facile synthesis processes and transfer-free film coating. Hybrid integrated devices with rGO can be readily fabricated by further reducing GO within the hybrid devices. Therefore, the GO fabrication techniques can be leveraged for large-scale manufacturing of hybrid integrated devices with rGO.

In **Fig. 3(c)**, slightly increased anisotropy ratio is also achieved for thicker GO films. For unreduced GO, the anisotropy ratios are ~4.4 and ~4.5 for the films including 1 and 2 layers of GO, respectively. For 1 layer of rGO at $T_R$ = ~100 °C, the anisotropy ratio is ~6, in contrast to ~7 for 2 layers of rGO at that same $T_R$. These results reflect that the thicker film exhibits more significant anisotropy in loss for both GO and rGO.

In **Fig. 3(d)**, we compare the measured *PDL* values with those obtained from optical mode simulations. We show the results at different $T_R$ for the hybrid devices with 1 and 2 layers of GO. In our simulations, we assumed that the GO films were isotropic with the same values of $k$ (*i.e.*, $k_{TE}$ in **Fig. 3(b)**) for both TE and TM polarizations. As a result, the simulated *PDL* values represent the polarization selectivity caused by the polarization-dependent mode overlap with the GO films, and the variation between the simulated and measured *PDL* values characterizes the extra polarization selectivity enabled by the loss anisotropy of 2D GO films. For all different $T_R$ and $N$, the simulated *PDL*'s exhibited positive values, reflecting that the polarization-dependent mode overlap with GO contributes to the overall *PDL*.

In **Fig. 3(e-i)** and **3(e-ii)**, we further calculated the fractional contributions to the overall *PDL* from the polarization-dependent mode overlap and the material loss anisotropy (where the sum of these two fractions equals 1). For all different $T_R$ and $N$, over 65% of the polarization selectivity is attributed to the loss anisotropy of 2D GO films. This highlights its dominance in enabling the functionality of the optical polarizer. It is also interesting to note that the fractional contribution from the loss anisotropy increases as $T_R$ increases. This is mainly due to the fact there is more significant loss anisotropy for rGO, as we discussed in **Fig. 3(c)**.

*Dependence on input power and wavelength*

For the experiments in **Fig. 2**, the *IL* was measured at a low input CW power of $P_{in}$ = ~0 dBm, to ensure that the GO films remained unaffected by photothermal reduction induced by the CW light. In **Fig. 4**, we further increase the input CW power $P_{in}$ to induce photothermal reduction of GO and characterize the changes in the polarization selectivity of the hybrid waveguides. Compared to GO reduction caused by heating the entire chip on a hot plate (as we did for the experiments in **Fig. 2**), using input CW power can trigger localized photothermal reduction of GO in the hybrid waveguides, along with dynamic changes in the GO film properties.

**Figure 4(a-i)** shows the measured *IL* versus $P_{in}$ for the hybrid waveguide with 1 layer of unreduced GO (*i.e.*, we directly measured the *IL* without heating the GO-coated chip on a hot plate, unlike what we did later in **Fig. 4(b)** and **4(c)**). The GO film length in the hybrid waveguide was $L_{GO}$ = ~0.4 mm, and the wavelength of the input CW light was ~1550 nm. We measured the *IL* for both TE and TM polarizations, and the results were recorded only when a steady thermal equilibrium state with stable output power was achieved. We chose an input power range of $P_{in} \leq$ ~25 dBm because the polymer layers in the self-assembled films cannot withstand input powers beyond this range.

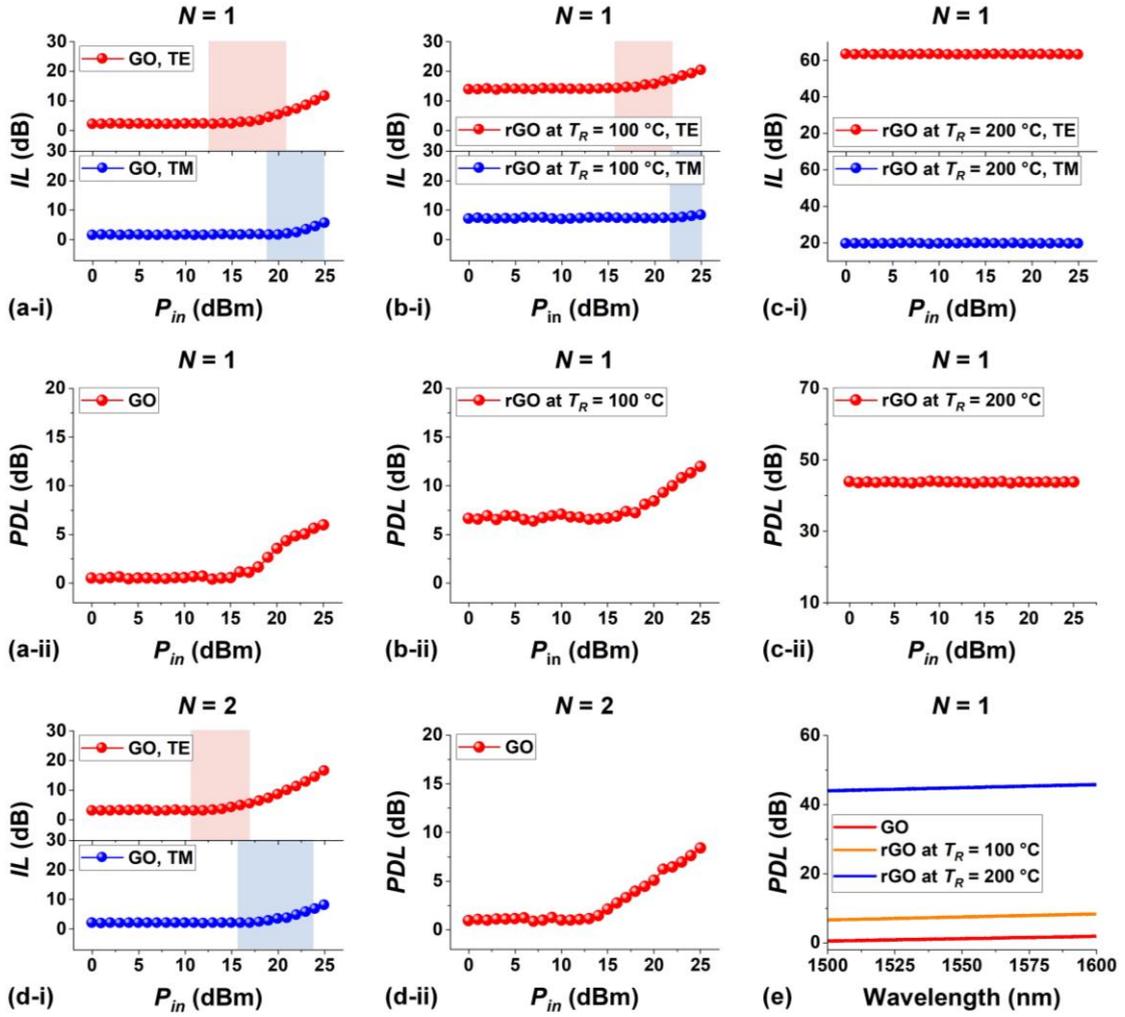

**Fig. 4** | (**a**) Measured (**i**) TE- and TM-polarized *IL* and (**ii**) calculated *PDL* versus input power ($P_{in}$) for the hybrid waveguide with 1 layer of GO. (**b**) – (**c**) Measured (**i**) TE- and TM-polarized *IL* and (**ii**) calculated *PDL* versus $P_{in}$ for the hybrid waveguide with 1 layer of rGO after heating at $T_R = \sim100$ °C and $\sim200$ °C, respectively. (**d**) Measured (**i**) TE- and TM-polarized *IL* and (**ii**) calculated *PDL* versus $P_{in}$ for the hybrid waveguide with 2 layers of GO. In (**a**) – (**d**), the red and blue shaded areas in (**i**) indicate the power ranges associated with reversible GO reduction for TE and TM polarizations, respectively. (**e**) Measured *PDL* versus input CW wavelength for the hybrid waveguide with 1 layer of unreduced GO, rGO at $T_R = \sim100$ °C, and rGO at $T_R = \sim200$ °C. In (**a**) – (**e**), the GO film length was $\sim0.4$ mm. In (**a**) – (**d**), the input CW wavelength was $\sim1550$ nm. In (**e**), the input CW power was $P_{in} = \sim0$ dBm.

In **Fig. 4(a-i)**, the TE-polarized *IL* remained constant at $\sim2$ dB when $P_{in} \leq \sim13$ dBm, indicating that the GO film was not reduced in this power range. For $P_{in} \geq \sim13$ dBm, the TE-polarized *IL* increased with $P_{in}$, and reached $\sim12$ dB at $P_{in} = \sim25$ dBm. This reflects that there was loss increase induced by photothermal reduction of GO at high light powers. We also note that the reduction of GO exhibited reversibility within a power range of $\sim13$ dBm $\leq P_{in} \leq \sim21$ dBm, as indicated by the red shaded area. In this power range, after turning off the high-power input and remeasuring the *IL* with a low input power of $P_{in}$ = ~0 dBm, the *IL* returned to ~2 dB (*i.e.*, the *IL* for unreduced GO when $P_{in} \leq$ ~13 dBm). This reversibility indicates that the photothermally reduced GO was unstable in nature, which reverted to the unreduced status after cooling down in an oxygen-containing ambient. As $P_{in}$ continued to rise above ~21 dBm, there was permanent increase in the *IL* after turning off the high-power input and remeasuring at $P_{in}$ = ~0 dBm. This reflects that there was permanent reduction of GO induced by the high CW power in this range, where the chemical bonds between the OFGs and the carbon network were irreversibly broken, resulting in a lasting alteration in GO's atomic structure and material properties. The photothermal reduction of GO in GO-Si waveguides is more significant as compared to that observed for GO-silicon nitride and GO-doped silica waveguides [31, 53], mainly due to the stronger GO mode overlap in the GO-Si waveguides.

In **Fig. 4(a-i)**, the TM-polarized *IL* increased when $P_{in} \geq$ ~19 dBm, reaching ~6 dB at $P_{in}$ = ~25 dBm. Compared to TE polarization, the power threshold for initiating photothermal reduction of GO was higher for TM polarization. This can be attributed to weaker photo-thermal effects for TM polarization that result from lower absorption for out-of-plane light waves in the anisotropic 2D GO films. **Figure 4(a-ii)** shows the corresponding *PDL* versus $P_{in}$ extracted from **Fig. 2(a-i)**. The *PDL* exhibited no significant changes when $P_{in} \leq$ ~13 dBm. However, when $P_{in}$ exceeded ~13 dBm, there was an obvious increase in the *PDL* as $P_{in}$ increased. This indicates that the polarization selectivity was enhanced by increasing the input power.

**Figures 4(b)** and **4(c)** show the corresponding results for the hybrid waveguides with 1 layer of rGO after heating at $T_R$ = ~100 °C and ~200 °C, respectively. For comparison, the GO film length was the same as that of the hybrid waveguide in **Fig. 4(a)**. Before measuring the *IL*, the GO-coated chip was heated on a hot plate for 15 minutes, as we did in **Fig. 2**. According to the results in **Fig. 2**, the GO films in the hybrid waveguides were reduced after heating at $T_R$ = ~100 °C and ~200 °C. For rGO at $T_R$ = ~100 °C in **Fig. 4(b)**, loss increase induced by photothermal reduction was observed for TE polarization when $P_{in} \geq$ ~16 dBm. The power threshold of ~16 dBm was higher than that for unreduced GO (*i.e.*, ~13 dBm in **Fig. 4(a)**). This indicates that unreduced GO was more easily reduced by the applied CW power, and higher power is required to trigger photothermal reduction of rGO.

For rGO at $T_R$ = ~200 °C in **Fig. 4(c)**, increasing $P_{in}$ did not result in any significant variations in the *IL* and *PDL*. These results further confirm that the photothermal reduction behaviour of GO becomes less obvious as the degree of reduction increases. According to Ref. [34], rGO exhibits higher thermal conductivity compared unreduced GO, and the thermal conductivity increases with the degree of reduction. The relatively high thermal conductivity of rGO leads to a lower heat accumulation efficiency, which in turn diminishes the photothermal effects and the power-dependent response. In addition to exhibiting a higher anisotropy ratio in **Fig. 3(c)**, rGO shows better thermal stability and stronger immunity to photothermal reduction than GO, making it attractive for implementing optical polarizers operating at high temperatures and input powers.

**Figure 4(d)** shows the corresponding results for the hybrid waveguide with 2 layers of unreduced GO (*i.e.*, without heating the GO-coated chip on a hot plate). Loss increase induced by photothermal reduction was observed for TE polarization when $P_{in} \geq$ ~11 dBm, and reversible GO reduction was observed when ~11 dBm $\leq P_{in} \leq$ ~17 dBm. Compared to the results in **Fig. 4(a)**, the hybrid waveguide with a thicker GO film exhibited a lower power threshold for initiating photothermal reduction and a smaller power range for reversible GO reduction. These reflect more significant photothermal effects in thicker GO films.

**Figure 4(e)** shows the measured *PDL* versus input CW wavelength for the hybrid waveguides with 1 layer of GO, rGO at $T_R$ = ~100 °C, and rGO at $T_R$ = ~200 °C. For comparison, the input CW power was kept the same as $P_{in}$ = ~0 dBm. For all three waveguides, the *PDL* exhibited a very small variation (< 1 dB) across the measured wavelength range of ~1500 – 1600 nm. This reflects the broad operation bandwidth for these waveguide polarizers. We also note that there was a slight increase in the *PDL* as the wavelength increased, mainly due to a minor change in the mode overlap with GO induced by dispersion.

*Polarization-selective microring resonators*

In addition to waveguide polarizers, we also investigate polarization-selective MRRs by integrating 2D GO films onto silicon MRRs. **Figure 5(a)** shows the schematic of a silicon MRR coated with 1 layer of GO, and a microscopic image of the fabricated device is provided in **Fig. 5(b)**. The silicon MRR had a radius of ~20 μm, and the length of the opened windows (*i.e.*, the length of the coated GO film) was ~10 μm. The ring and the bus waveguide in the MRR had the same waveguide cross-section of ~400 nm × 220 nm – identical to that for the waveguide polarizers in **Fig. 2**. The hybrid MRR and the hybrid waveguides in **Fig. 2** were fabricated on the same SOI chip via the same processes.

In **Fig. 5(c)**, we compare the TE- and TM- polarized transmission spectra for the hybrid MRRs with 1 layer of GO at different degrees of reduction. All the spectra were measured by scanning the wavelength of an input CW light with a power of $P_{in}$ = ~-10 dBm (which did not induce any significant photo-thermal effects in the GO films). We first measured the device with unreduced GO in **Fig. 5(b)** before heating it on a hot plate (the results are labeled as 'initial'). Then, we measured the same device after heating it on a hot plate at various temperatures $T_R$ ranging from ~50 to ~200 °C (for 15 minutes, as we did in **Fig. 2**).

**Figure 5(d)** shows the extinction ratios (*ER*'s) of the hybrid MRRs extracted from **Fig. 5(c)**. As can be seen, the *ER* of the hybrid MRR after heating at $T_R$ = ~50 °C showed negligible difference as compared to that of the unheated MRR. This shows agreement with the results in **Fig. 2** and provides further evidence that the reduction of GO did not occur at $T_R$ = ~50 °C. For $T_R \geq 100$ °C, an increase in the *ER* was observed as $T_R$ increased, particularly for TE polarization. This was because the uncoated silicon MRR we chose was over-coupled [54, 55]. As $T_R$ increased, the degree of reduction for GO also increased, leading to higher loss of the GO film. As the loss induced by GO increased, the difference between the round-trip loss and the coupling strength in the hybrid MRR became smaller, resulting in a higher *ER* (*i.e.*, more approaching the critical coupling condition [56]).

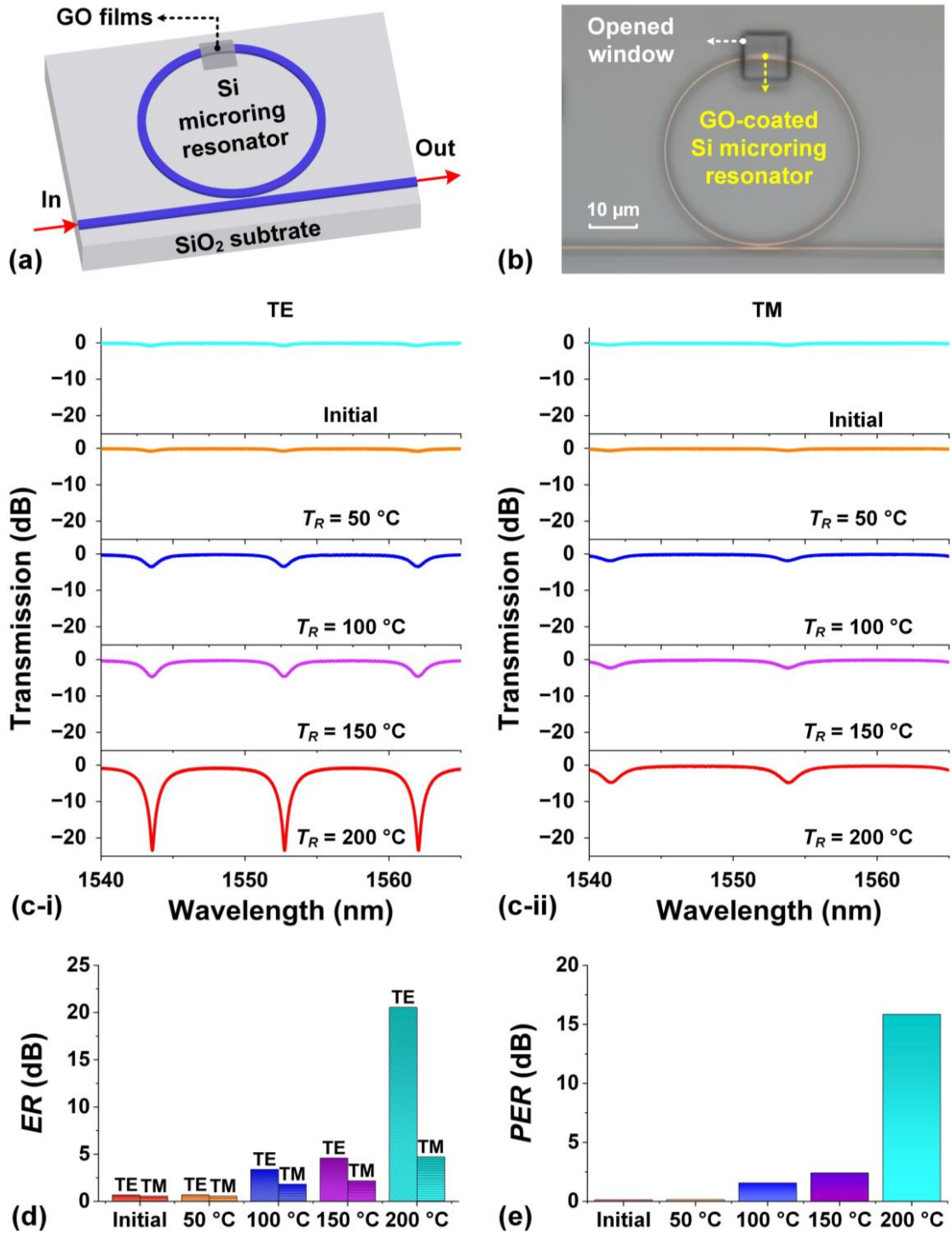

**Fig. 5** | (**a**) Schematic illustration of a GO-coated silicon microring resonator (MRR) as a polarization-selective MRR. (**b**) Microscopic image of a fabricated silicon MRR coated with 1 layer of unreduced GO. (**c**) Measured (**i**) TE- and (**ii**) TM-polarized transmission spectra of the hybrid MRR with 1 layer of GO at different degrees of reduction. The same hybrid MRR underwent heating at temperatures $T_R$ ranging from ~50 to 200 °C prior to the measurement. The corresponding results measured at room temperature before heating (initial) are also shown for comparison. (**d**) Extinction ratios (*ER*'s) for the MRRs extracted from (c). (**e**) Polarization extinction ratios (*PER*'s) extracted from (d). In (**c**) – (**e**), the CW input power was $P_{in}$ = ~-10 dBm.

**Figure 5(e)** shows the *PER* obtained by subtracting the TM-polarized *ER* from the TE-polarized *ER* in **Fig. 5(d)**. As can be seen, the hybrid MRR with unreduced GO exhibited a low polarization selectivity, with its *PER* being less than ~1 dB. In contrast, The *PER* increased with $T_R$ when $T_R \geq 100$ °C. After heating at $T_R =$ ~200 °C, the hybrid MRR exhibited a high *PER* of ~16 dB, highlighting its excellent polarization selectivity.

In **Fig. 6**, we characterize the power-dependent response for the polarization-selective MRRs in **Fig. 5** by increasing the input CW power to induce photothermal reduction of GO. In **Fig. 5**, we measured the MRRs' transmission spectra by using a single input CW light with a low power of $P_{in} =$ ~-10 dBm. In **Fig. 6**, we employed two CW inputs in our measurements. The first one with a power of $P_p$ was employed as a pump injecting into one of the MRR's resonances near ~1550 nm. The wavelength of this input CW light was slightly tuned around the resonance until a steady thermal equilibrium state with stable output power was achieved. After this, the second CW light, with a power of ~-10 dBm (*i.e.*, the same as that in **Fig. 5**), was employed as a low-power probe to scan the MRR's transmission spectrum. Compared to directly using a high-power CW light to scan the spectrum, this approach would not induce significant asymmetry in the measured resonance spectral lineshape caused by optical bistability [57, 58], thus allowing for a higher accuracy in characterizing the MRR's extinction ratio. In our measurements, the CW pump power $P_p$ was ≤ ~15 dBm to prevent damages to the polymer layers in the self-assembled films. We also chose $P_p \geq$ ~0 dBm to ensure that the power of the probe light remained negligible compared to $P_p$.

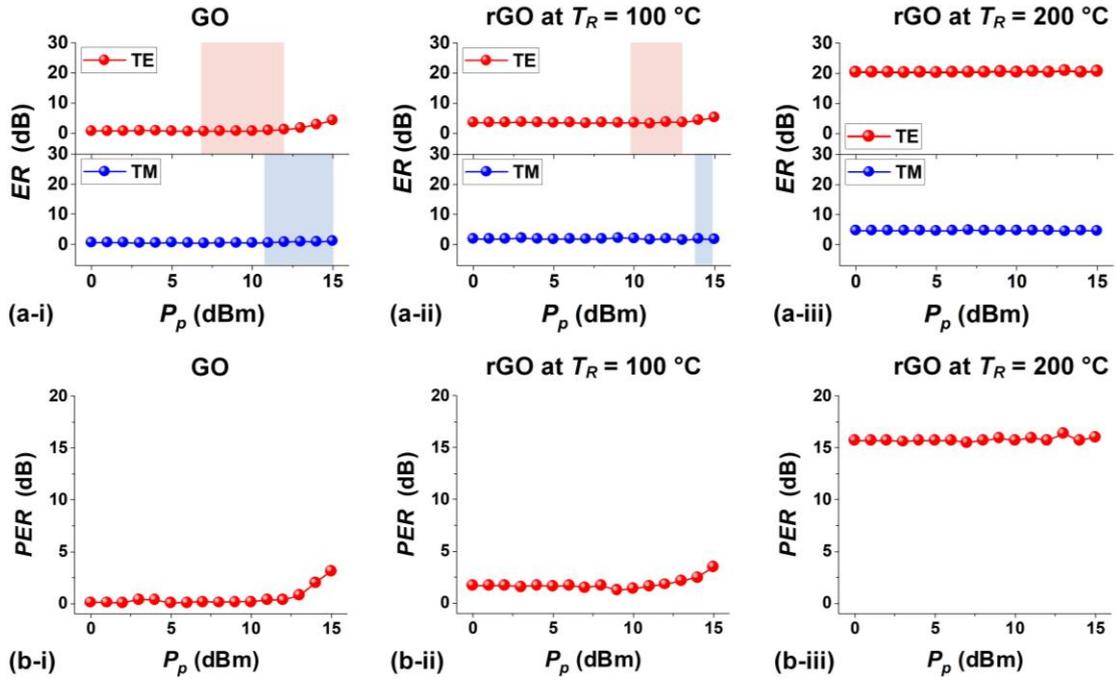

**Fig. 6** | (**a**) Measured TE- and TM-polarized *ER* versus input CW pump power $P_p$ for the hybrid MRR with 1 layer of GO at different degrees of reduction. (**i**) – (**iii**) show the results measured for the same hybrid MRR with 1 layer of GO, rGO after heating at $T_R$ = ~100 °C, and rGO after heating at ~200 °C, respectively. (**b**) *PER*'s extracted from (a). In (**a**) – (**b**), the red and blue shaded areas indicate the power ranges associated with reversible GO reduction for TE and TM polarizations, respectively.

In **Fig. 6(a)**, we plot TE- and TM-polarized *ER* versus input CW pump power $P_p$. We first measured a hybrid MRR with unreduced GO. As shown in **Fig. 6(a-i)**, the TE-polarized *ER* exhibited no significant variations when $P_p \leq$ ~7 dBm. When $P_p \geq$ ~7 dBm, it increased with $P_p$, indicating that there was increased loss induced by localized photothermal reduction of GO. The power threshold of ~7 dBm for the hybrid MRR was much lower than that for a comparable hybrid waveguide (*i.e.*, ~13 dBm in **Fig. 4(a)**), reflecting more significant photothermal effects in the hybrid MRR enabled by the resonance enhancement effect. Compared to TE polarization, a higher power threshold of ~11 dB was observed for TM polarization, further indicating the anisotropy of the 2D GO film. For both polarizations, reversible GO reduction behaviour was also observed within specific power ranges – similar to the results in **Fig. 4(a)**.

**Figures 6(a-ii)** and **6(a-iii)** show the corresponding results for the hybrid MRR with 1 layer of rGO after heating at $T_R = $ ~100 °C and ~200 °C, respectively. For the device with rGO at $T_R = $ ~100 °C, the increase in *ER* caused by localized photothermal reduction of GO was observed when $P_p \geq$ ~10 dBm for TE polarization and $P_p \geq$ ~14 dBm for TM polarization. These power thresholds are higher than those in **Fig. 6(a)** for the device with unreduced GO, further confirming that a higher CW power is needed to induce photothermal reduction of rGO. For the device with rGO at $T_R = $ ~200 °C, no significant variations in the *ER* were observed for both polarizations within the measured input pump power range. This highlights that the highly reduced GO exhibited even less noticeable photothermal reduction behaviour, showing agreement with the results in **Fig. 4c**.

**Figure 6(b)** shows the *PER* calculated from **Fig. 6(a)**. In **Fig. 6(b-i)**, the hybrid MRR with unreduced GO exhibited a low *PER* < ~1 dB when $P_p <$ ~7 dBm, and the *PER* increased when $P_p \geq$ ~7 dBm, reaching ~3 dB at $P_{in} = $ ~15 dBm. For the hybrid MRR with rGO at $T_R = $ ~100 °C, the *PER* increased from ~2 dB in the low-power state without photothermal reduction to ~4 dB at $P_p = $ ~15 dBm. For rGO at $T_R = $ ~200 °C, the *PER* remained unchanged at ~16 dB as $P_p$ increased from ~0 dBm to ~15 dBm. These results further confirms that the hybrid MRR with highly reduced GO is less susceptible to variations in the input power and shows a better power stability.

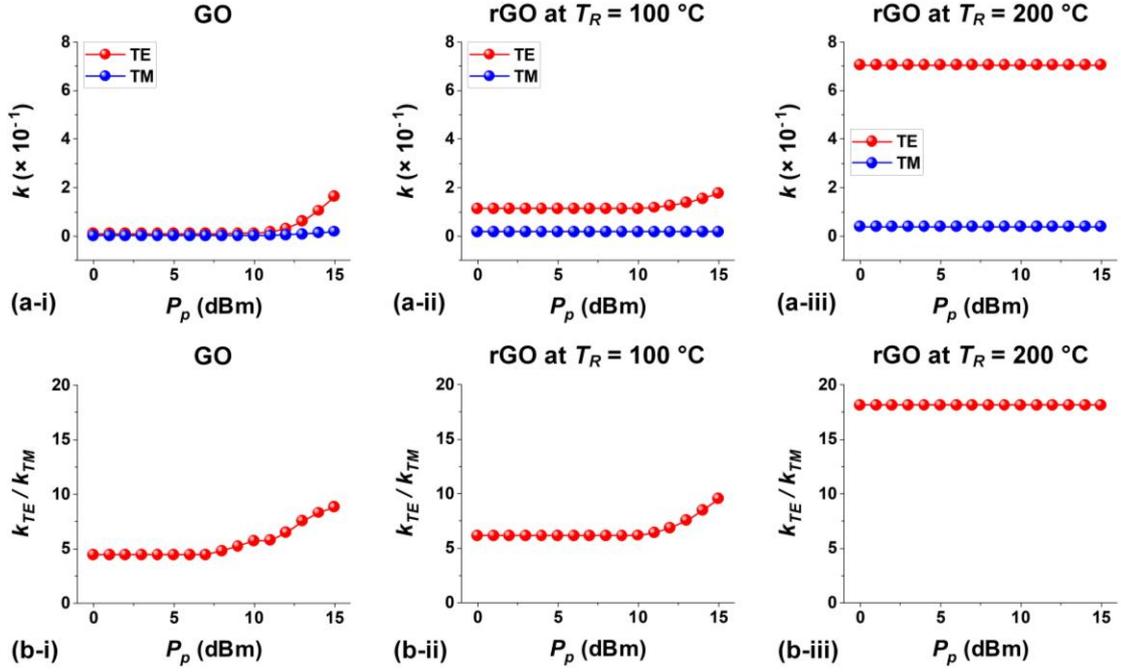

**Fig. 7** | (**a**) Extinction coefficients ($k$'s) of 2D GO films versus $P_p$ obtained by fitting the results in **Fig. 6(a)** with optical mode simulations. (**i**) – (**iii**) show the results for GO, rGO after heating at $T_R$ = ~100 °C, and rGO after heating at ~200 °C, respectively. (**b**) Anisotropy ratios of $k$ values for TE and TM polarizations ($k_{TE}$ / $k_{TM}$) extracted from (**a**).

In **Fig. 6**, the variations in the *ER* of the hybrid MRRs cannot directly indicate changes in the properties of the GO films. To address this, we further extracted the extinction coefficients ($k$'s) of 2D GO films by fitting the results in **Fig. 6(a)** with theory and plotted them in **Fig. 7(a)**. In our fitting process, we first obtained the GO-induced *EPL* by fitting the measured transmission spectrum of the hybrid MRR based on the scattering matrix method [59]. After that, the $k$ of 2D GO film was extracted from the obtained *EPL* by using the same method as we used in **Fig. 3(b)**. Note that the photothermal changes in GO films coated on integrated waveguides or MRRs actually exhibit nonuniform behavior along the direction of light propagation [33]. This occurs because, as the light power diminishes along the 2D film, the photothermal effects become weaker, resulting in a smaller difference in properties between the photothermally reduced GO and the unreduced GO. For simplification, in our fitting process we regarded the 10-μm-long GO or rGO films in the hybrid MRRs as uniform films with consistent loss. In principle, this approximation can lead to slight deviations in the fit $k$ values, particularly at a high $P_p$. Despite this, the fit $k$ can still be regarded as an average value reflecting the over-all loss performance of the GO films at different $P_p$.

For unreduced GO in **Fig. 7(a-i)**, the $k$ values at low pump powers (*e.g.*, $P_p = \sim 0$ dBm) are ~0.0088 and ~0.0017 for TE and TM polarizations, respectively. These values obtained from the MRR experiment show good agreement with those obtained from the waveguide experiment in **Fig. 3(b)**, reflecting the consistency of our GO film fabrication process. The $k$ for TE polarization increases when $P_p \geq \sim 7$ dBm and reaches ~0.1634 at $P_{in} = \sim 15$ dBm – ~17 times of the $k$ at $P_p < \sim 7$ dBm. This suggests that the change in $k$ induced by localized photothermal reduction of GO is quite significant, even though the variation in *ER* shown in **Fig. 6(a)** is not very noticeable. This is mainly due to the fact the *ER* in **Fig. 6(a)** was plotted on a dB scale, which results in less significant change for the *ER* with a lower value.

**Figures 7(a-ii)** and **7(a-iii)** show the corresponding results for the hybrid MRR with 1 layer of rGO after heating at $T_R = \sim 100$ °C and ~200 °C, respectively. For rGO at $T_R = \sim 100$ °C, the TE-polarized $k$ increases from ~0.1013 at $P_p < \sim 10$ dBm to ~0.1670 at $P_p = \sim 15$ dBm. Whereas the $k$ for TM polarization slightly increases from ~0.0183 at $P_p < \sim 15$ dBm to ~0.0197 at $P_p = \sim 15$ dBm. In contrast, the $k$ of rGO at $T_R = \sim 200$ °C remains constant for both polarizations (*i.e.*, $k = \sim 0.7022$ for TE polarization and $k = \sim 0.0367$ for TM polarization) as $P_{in}$ increases from ~0 dBm to ~15 dBm.

**Figure 7(b)** shows the anisotropy ratios calculated from **Fig. 7(a)**. In **Fig. 7(b-i)**, the anisotropy ratio for unreduced GO remains constant at ~4.5 when $P_p < 7$ dBm, showing agreement with the results in **Fig. 3(c)**. For $P_p \geq 7$ dBm, the anisotropy ratio increases with $P_p$, achieving a maximum value of ~8.8 at $P_{in}$ = ~15 dBm. For rGO at $T_R$ = ~100 °C, the anisotropy ratio remains unchanged at ~6.2 when $P_p$ < ~10 dBm before experiencing a gradual increase to ~9.6 at $P_p$ = ~15 dBm. For rGO at $T_R$ = ~200 °C, the anisotropy ratio remains unchanged at ~18.1 within the measured input pump power range. These results further confirm that 2D GO films exhibit more significant loss anisotropy as the degree of reduction increases. This will work be aided with the use of other novel 2D materials [60 – 81] that will be extremely useful for all forms of microcombs [82 – 166] with the use of novel designs [167 – 174] for a wide range of applications.

**Conclusions**

In summary, we experimentally demonstrate integrated waveguide and MRR polarizers incorporating rGO. We integrate 2D GO films onto silicon photonic devices with precise control over their thicknesses and sizes, and use two methods – uniform thermal reduction and localized photothermal reduction – to reduce the GO films. Detailed measurements are performed for devices with different lengths, thicknesses, and reduction levels of the GO films. The results show that the devices with rGO exhibit better polarizer performance than those with GO. A maximum *PDL* of ~47 dB is achieved for the hybrid waveguide with rGO, and the hybrid MRR with rGO achieves a maximum *PER* of ~16. By fitting the experimental results with theory, it reveals that rGO exhibits more significant loss anisotropy, with an anisotropy ratio more than 4 times that of GO. In addition, rGO also exhibits enhanced thermal stability and lower sensitivity to photothermal reduction. Our work opens up new opportunities for implementing high-performance polarization-selective devices through on-chip integration of 2D rGO films.

## Competing interests

The authors declare no competing financial interests.

## References

[1] Q. Bao, H. Zhang, B. Wang, Z. Ni, C. H. Y. X. Lim, Y. Wang, D. Y. Tang, and K. P. Loh, "Broadband graphene polarizer," *Nature Photonics,* vol. 5, no. 7, pp. 411-415, 2011/07/01, 2011.

[2] Y. Yan, G. Xie, M. P. J. Lavery, H. Huang, N. Ahmed, C. Bao, Y. Ren, Y. Cao, L. Li, Z. Zhao, A. F. Molisch, M. Tur, M. J. Padgett, and A. E. Willner, "High-capacity millimetre-wave communications with orbital angular momentum multiplexing," *Nature Communications,* vol. 5, no. 1, pp. 4876, 2014/09/16, 2014.

[3] C. He, H. He, J. Chang, B. Chen, H. Ma, and M. J. Booth, "Polarisation optics for biomedical and clinical applications: a review," *Light: Science & Applications,* vol. 10, no. 1, pp. 194, 2021/09/22, 2021.

[4] D. Dai, L. Liu, S. Gao, D.-X. Xu, and S. He, "Polarization management for silicon photonic integrated circuits," *Laser & Photonics Reviews,* vol. 7, no. 3, pp. 303-328, 2013.

[5] K. Serkowski, D. S. Mathewson, and V. L. Ford, "Wavelength dependence of interstellar polarization and ratio of total to selective extinction," *The Astrophysical Journal,* vol. 196, pp. 261-290, February 01, 1975, 1975.

[6] T. J. Wang, Q. Y. He, J. Y. Gao, Y. Jiang, Z. H. Kang, H. Sun, L. S. Yu, X. F. Yuan, and J. Wu, "Efficient electrooptically Q-switched Er:Cr:YSGG laser oscillator-amplifier system with a Glan-Taylor prism polarizer," *Laser Physics,* vol. 16, no. 12, pp. 1605-1609, 2006/12/01, 2006.

[7] A. Rahnama, T. Dadalyan, K. Mahmoud Aghdami, T. Galstian, and P. R. Herman, "In-Fiber Switchable Polarization Filter Based on Liquid Crystal Filled Hollow-Filament Bragg Gratings," *Advanced Optical Materials,* vol. 9, no. 19, pp. 2100054, 2021.

[8] E. Saitoh, Y. Kawaguchi, K. Saitoh, and M. Koshiba, "TE/TM-Pass Polarizer Based on Lithium Niobate on Insulator Ridge Waveguide," *IEEE Photonics Journal,* vol. 5, no. 2, pp. 6600610-6600610, 2013.

[9] J. Wang, J.-Y. Yang, I. M. Fazal, N. Ahmed, Y. Yan, H. Huang, Y. Ren, Y. Yue, S. Dolinar, M. Tur, and A. E. Willner, "Terabit free-space data transmission employing orbital angular momentum multiplexing," *Nature Photonics,* vol. 6, no. 7, pp. 488-496, 2012/07/01, 2012.

[10] N. Bozinovic, Y. Yue, Y. Ren, M. Tur, P. Kristensen, H. Huang, A. E. Willner, and S. Ramachandran, "Terabit-Scale Orbital Angular Momentum Mode Division Multiplexing in Fibers," *Science,* vol. 340, no. 6140, pp. 1545-1548, 2013.

[11] X. Guan, P. Chen, S. Chen, P. Xu, Y. Shi, and D. Dai, "Low-loss ultracompact transverse-magnetic-pass polarizer with a silicon subwavelength grating waveguide," *Optics Letters,* vol. 39, no. 15, pp. 4514-4517, 2014/08/01, 2014.

[12] Y. Huang, S. Zhu, H. Zhang, T.-Y. Liow, and G.-Q. Lo, "CMOS compatible horizontal nanoplasmonic slot waveguides TE-pass polarizer on silicon-on-insulator platform," *Optics Express,* vol. 21, no. 10, pp. 12790-12796, 2013/05/20, 2013.

[13] M. Z. Alam, J. S. Aitchison, and M. Mojahedi, "Compact and silicon-on-insulator-compatible hybrid plasmonic TE-pass polarizer," *Optics Letters,* vol. 37, no. 1, pp. 55-57, 2012/01/01, 2012.

[14] Y. Zhang, J. Wu, L. Jia, D. Jin, B. Jia, X. Hu, D. J. Moss, and Q. Gong, "Advanced optical polarizers based on 2D materials," *npj Nanophotonics,* vol. 1, no. 1, pp. 28, 2024/07/17, 2024.

[15] H. Lin, Y. Song, Y. Huang, D. Kita, S. Deckoff-Jones, K. Wang, L. Li, J. Li, H. Zheng, Z. Luo, H. Wang, S. Novak, A. Yadav, C.-C. Huang, R.-J. Shiue, D. Englund, T. Gu, D. Hewak, K. Richardson, J. Kong, and J. Hu, "Chalcogenide glass-on-graphene photonics," *Nature Photonics,* vol. 11, no. 12, pp. 798-805, 2017/12/01, 2017.

[16] J. Guo, Y. Liu, L. Lin, S. Li, J. Cai, J. Chen, W. Huang, Y. Lin, and J. Xu, "Chromatic Plasmonic Polarizer-Based Synapse for All-Optical Convolutional Neural Network," *Nano Letters,* vol. 23, no. 20, pp. 9651-9656, 2023/10/25, 2023.

[17] S. Wang, S. Wen, Z.-L. Deng, X. Li, and Y. Yang, "Metasurface-Based Solid Poincaré Sphere Polarizer," *Physical Review Letters,* vol. 130, no. 12, pp. 123801, 03/23/, 2023.

[18] K. S. Novoselov, "Electric field effect in atomically thin carbon films," *Science,* vol. 306, pp. 666, 2004.

[19] M. Zeng, Y. Xiao, J. Liu, K. Yang, and L. Fu, "Exploring Two-Dimensional Materials toward the Next-Generation Circuits: From Monomer Design to Assembly Control," *Chemical Reviews,* vol. 118, no. 13, pp. 6236-6296, 2018/07/11, 2018.

[20] F. Xia, H. Wang, D. Xiao, M. Dubey, and A. Ramasubramaniam, "Two-dimensional material nanophotonics," *Nature Photonics,* vol. 8, no. 12, pp. 899-907, 2014/12/01, 2014.

[21] K. S. Novoselov, A. Mishchenko, A. Carvalho, and A. H. Castro Neto, "2D materials and van der Waals heterostructures," *Science,* vol. 353, no. 6298, pp. aac9439, 2016.

[22] J. T. Kim, and H. Choi, "Polarization Control in Graphene-Based Polymer Waveguide Polarizer," *Laser & Photonics Reviews,* vol. 12, no. 10, pp. 1800142, 2018.

[23] J.-l. Kou, J.-h. Chen, Y. Chen, F. Xu, and Y.-q. Lu, "Platform for enhanced light−graphene interaction length and miniaturizing fiber stereo devices," *Optica,* vol. 1, no. 5, pp. 307-310, 2014/11/20, 2014.

[24] J. Wu, Y. Yang, Y. Qu, X. Xu, Y. Liang, S. T. Chu, B. E. Little, R. Morandotti, B. Jia, and D. J. Moss, "Graphene Oxide Waveguide and Micro-Ring Resonator Polarizers," *Laser & Photonics Reviews,* vol. 13, no. 9, pp. 1900056, 2019.

[25] D. Jin, J. Wu, J. Hu, W. Liu, Y. Zhang, Y. Yang, L. Jia, D. Huang, B. Jia, and D. J. Moss, "Silicon photonic waveguide and microring resonator polarizers

incorporating 2D graphene oxide films," *Applied Physics Letters,* vol. 125, no. 5, 2024.

[26] W. H. Lim, Y. K. Yap, W. Y. Chong, C. H. Pua, N. M. Huang, R. M. De La Rue, and H. Ahmad, "Graphene oxide-based waveguide polariser: From thin film to quasi-bulk," *Optics Express,* vol. 22, no. 9, pp. 11090-11098, 2014/05/05, 2014.

[27] L. Zhuo, D. Li, W. Chen, Y. Zhang, W. Zhang, Z. Lin, H. Zheng, W. Zhu, Y. Zhong, J. Tang, G. Lu, W. Fang, J. Yu, and Z. Chen, "High performance multifunction-in-one optoelectronic device by integrating graphene/MoS2 heterostructures on side-polished fiber," *Nanophotonics,* vol. 11, no. 6, pp. 1137-1147, 2022.

[28] Y. Tan, R. He, C. Cheng, D. Wang, Y. Chen, and F. Chen, "Polarization-dependent optical absorption of MoS2 for refractive index sensing," *Scientific Reports,* vol. 4, no. 1, pp. 7523, 2014/12/17, 2014.

[29] N. Berahim, I. S. Amiri, T. Anwar, S. R. Azzuhri, M. N. S. Mohd Nasir, R. Zakaria, W. Y. Chong, C. K. Lai, S. H. Lee, H. Ahmad, M. A. Ismail, and P. Yupapin, "Polarizing effect of MoSe2-coated optical waveguides," *Results in Physics,* vol. 12, pp. 7-11, 2019/03/01/, 2019.

[30] Y. Yang, Y. Zhang, J. Zhang, X. Zheng, Z. Gan, H. Lin, M. Hong, and B. Jia, "Graphene Metamaterial 3D Conformal Coating for Enhanced Light Harvesting," *ACS Nano,* vol. 17, no. 3, pp. 2611-2619, 2023/02/14, 2023.

[31] J. Wu, Y. Yang, Y. Qu, L. Jia, Y. Zhang, X. Xu, S. T. Chu, B. E. Little, R. Morandotti, B. Jia, and D. J. Moss, "2D Layered Graphene Oxide Films Integrated with Micro-Ring Resonators for Enhanced Nonlinear Optics," *Small,* vol. 16, no. 16, pp. 1906563, 2020.

[32] K.-T. Lin, H. Lin, T. Yang, and B. Jia, "Structured graphene metamaterial selective absorbers for high efficiency and omnidirectional solar thermal energy conversion," *Nature communications,* vol. 11, no. 1, pp. 1389, 2020.

[33] J. Wu, Y. Zhang, J. Hu, Y. Yang, D. Jin, W. Liu, D. Huang, B. Jia, and D. J. Moss, "2D graphene oxide films expand functionality of photonic chips," *Advanced Materials*, 2024.

[34] J. Hu, J. Wu, W. Liu, D. Jin, H. E. Dirani, S. Kerdiles, C. Sciancalepore, P. Demongodin, C. Grillet, C. Monat, D. Huang, B. Jia, and D. J. Moss, "2D Graphene Oxide: A Versatile Thermo-Optic Material," *Advanced Functional Materials,* vol. n/a, no. n/a, pp. 2406799, 2024.

[35] Y. Zhang, J. Wu, L. Jia, Y. Qu, Y. Yang, B. Jia, and D. J. Moss, "Graphene Oxide for Nonlinear Integrated Photonics," *Laser & Photonics Reviews,* vol. 17, no. 3, pp. 2200512, 2023/03/01, 2023.

[36] K. P. Loh, Q. Bao, G. Eda, and M. Chhowalla, "Graphene oxide as a chemically tunable platform for optical applications," *Nature Chemistry,* vol. 2, no. 12, pp. 1015-1024, 2010/12/01, 2010.

[37] G.-K. Lim, Z.-L. Chen, J. Clark, R. G. S. Goh, W.-H. Ng, H.-W. Tan, R. H. Friend, P. K. H. Ho, and L.-L. Chua, "Giant broadband nonlinear optical absorption response in dispersed graphene single sheets," *Nature Photonics,* vol. 5, no. 9, pp. 554-560, 2011/09/01, 2011.

[38] Z. Luo, P. M. Vora, E. J. Mele, A. T. C. Johnson, and J. M. Kikkawa, "Photoluminescence and band gap modulation in graphene oxide," *Applied Physics Letters,* vol. 94, no. 11, 2009.

[39] M. Fatkullin, D. Cheshev, A. Averkiev, A. Gorbunova, G. Murastov, J. Liu, P. Postnikov, C. Cheng, R. D. Rodriguez, and E. Sheremet, "Photochemistry dominates over photothermal effects in the laser-induced reduction of graphene oxide by visible light," *Nature Communications,* vol. 15, no. 1, pp. 9711, 2024/11/09, 2024.

[40] A. Bagri, C. Mattevi, M. Acik, Y. J. Chabal, M. Chhowalla, and V. B. Shenoy, "Structural evolution during the reduction of chemically derived graphene oxide," *Nature Chemistry,* vol. 2, no. 7, pp. 581-587, 2010/07/01, 2010.

[41] K. K. H. De Silva, H.-H. Huang, R. Joshi, and M. Yoshimura, "Restoration of the graphitic structure by defect repair during the thermal reduction of graphene oxide," *Carbon,* vol. 166, pp. 74-90, 2020/09/30/, 2020.

[42] Y. Zhu, S. Murali, W. Cai, X. Li, J. W. Suk, J. R. Potts, and R. S. Ruoff, "Graphene and Graphene Oxide: Synthesis, Properties, and Applications," *Advanced Materials,* vol. 22, no. 35, pp. 3906-3924, 2010.

[43] N. Ghofraniha, and C. Conti, "Graphene oxide photonics," *Journal of Optics,* vol. 21, no. 5, pp. 053001, 2019/04/09, 2019.

[44] Z. Wei, D. Wang, S. Kim, S.-Y. Kim, Y. Hu, M. K. Yakes, A. R. Laracuente, Z. Dai, S. R. Marder, C. Berger, W. P. King, W. A. de Heer, P. E. Sheehan, and E. Riedo, "Nanoscale Tunable Reduction of Graphene Oxide for Graphene Electronics," *Science,* vol. 328, no. 5984, pp. 1373-1376, 2010.

[45] D. Voiry, J. Yang, J. Kupferberg, R. Fullon, C. Lee, H. Y. Jeong, H. S. Shin, and M. Chhowalla, "High-quality graphene via microwave reduction of solution-exfoliated graphene oxide," *Science,* vol. 353, no. 6306, pp. 1413-1416, 2016.

[46] L. Liu, K. Xu, X. Wan, J. Xu, C. Y. Wong, and H. K. Tsang, "Enhanced optical Kerr nonlinearity of MoS2 on silicon waveguides," *Photonics Research,* vol. 3, no. 5, pp. 206-209, 2015/10/01, 2015.

[47] Q. Feng, H. Cong, B. Zhang, W. Wei, Y. Liang, S. Fang, T. Wang, and J. Zhang, "Enhanced optical Kerr nonlinearity of graphene/Si hybrid waveguide," *Applied Physics Letters,* vol. 114, no. 7, 2019.

[48] Y. Zhang, J. Wu, Y. Yang, Y. Qu, L. Jia, T. Moein, B. Jia, and D. J. Moss, "Enhanced Kerr Nonlinearity and Nonlinear Figure of Merit in Silicon Nanowires Integrated with 2D Graphene Oxide Films," *ACS Applied Materials & Interfaces,* vol. 12, no. 29, pp. 33094-33103, 2020/07/22, 2020.

[50] Y. Zhang, J. Wu, Y. Yang, Y. Qu, H. E. Dirani, R. Crochemore, C. Sciancalepore, P. Demongodin, C. Grillet, C. Monat, B. Jia, and D. J. Moss, "Enhanced self-phase modulation in silicon nitride waveguides integrated with 2D graphene oxide films," *IEEE Journal of Selected Topics in Quantum Electronics*, pp. 1-1, 2022.

[51] H. Cai, Y. Cheng, H. Zhang, Q. Huang, J. Xia, R. Barille, and Y. Wang, "Enhanced linear absorption coefficient of in-plane monolayer graphene on a silicon

microring resonator," *Optics Express,* vol. 24, no. 21, pp. 24105-24116, 2016/10/17, 2016.

[52] H. Li, Y. Anugrah, S. J. Koester, and M. Li, "Optical absorption in graphene integrated on silicon waveguides," *Applied Physics Letters,* vol. 101, no. 11, 2012.

[53] Y. Qu, J. Wu, Y. Yang, Y. Zhang, Y. Liang, H. El Dirani, R. Crochemore, P. Demongodin, C. Sciancalepore, C. Grillet, C. Monat, B. Jia, and D. J. Moss, "Enhanced Four-Wave Mixing in Silicon Nitride Waveguides Integrated with 2D Layered Graphene Oxide Films," *Advanced Optical Materials,* vol. 8, no. 23, pp. 2001048, 2020.

[54] J. Wu, P. Cao, X. Hu, T. Wang, M. Xu, X. Jiang, F. Li, L. Zhou, and Y. Su, "Nested Configuration of Silicon Microring Resonator With Multiple Coupling Regimes," *IEEE Photonics Technology Letters,* vol. 25, no. 6, pp. 580-583, 2013.

[55] W. Bogaerts, P. De Heyn, T. Van Vaerenbergh, K. De Vos, S. Kumar Selvaraja, T. Claes, P. Dumon, P. Bienstman, D. Van Thourhout, and R. Baets, "Silicon microring resonators," *Laser & Photonics Reviews,* vol. 6, no. 1, pp. 47-73, 2012/01/02, 2012.

[56] S. Feng, T. Lei, H. Chen, H. Cai, X. Luo, and A. W. Poon, "Silicon photonics: from a microresonator perspective," *Laser & Photonics Reviews,* vol. 6, no. 2, pp. 145-177, 2012.

[57] P. W. Smith, and W. J. Tomlinson, "Bistable optical devices promise subpicosecond switching," *IEEE Spectrum,* vol. 18, pp. 26-33, June 01, 1981, 1981.

[58] J. Hu, J. Wu, D. Jin, S. T. Chu, B. E. Little, D. Huang, R. Morandotti, and D. J. Moss, "Thermo-Optic Response and Optical Bistablility of Integrated High-Index Doped Silica Ring Resonators," *Sensors,* vol. 23, no. 24, pp. 9767, 2023.

[59] H. Arianfard, S. Juodkazis, D. J. Moss, and J. Wu, "Sagnac interference in integrated photonics," *Applied Physics Reviews,* vol. 10, no. 1, 2023.

60. Yuning Zhang, Jiayang Wu, Yunyi Yang, Yang Qu, Linnan Jia, Houssein El Dirani, Sébastien Kerdiles, Corrado Sciancalepore, Pierre Demongodin, Christian Grillet, Christelle Monat, Baohua Jia, and David J. Moss, "Enhanced supercontinuum generated in SiN waveguides coated with GO films", **Advanced Materials Technologies 8** (1) 2201796 (2023). DOI:10.1002/admt.202201796.

61. Jiayang Wu, H.Lin, David J. Moss, T.K. Loh, Baohua Jia, "Graphene oxide for electronics, photonics, and optoelectronics", ***Nature Reviews Chemistry* 7** (3) 162–183 (2023). doi.org/10.1038/s41570-022-00458-7.

62. Yang Qu, Jiayang Wu, Yuning Zhang, Yunyi Yang, Linnan Jia, Baohua Jia, and David J. Moss, "Photo thermal tuning in GO-coated integrated waveguides", Micromachines Vol. 13 1194 (2022). doi.org/10.3390/mi13081194

63. Yuning Zhang, Jiayang Wu, Yunyi Yang, Yang Qu, Houssein El Dirani, Romain Crochemore, Corrado Sciancalepore, Pierre Demongodin, Christian Grillet, Christelle Monat, Baohua Jia, and David J. Moss, "Enhanced self-phase modulation in silicon nitride waveguides integrated with 2D graphene oxide films", IEEE Journal of Selected Topics in Quantum Electronics Vol. 29 (1) 5100413 (2023). DOI: 10.1109/JSTQE.2022.3177385

64. Yuning Zhang, Jiayang Wu, Yunyi Yang, Yang Qu, Linnan Jia, Baohua Jia, and David J. Moss, "Enhanced spectral broadening of femtosecond optical pulses in silicon nanowires integrated with 2D graphene oxide films", Micromachines Vol. 13 756 (2022). DOI:10.3390/mi13050756.
65. Jiayang Wu, Yuning Zhang, Junkai Hu, Yunyi Yang, Di Jin, Wenbo Liu, Duan Huang, Baohua Jia, David J. Moss, "Novel functionality with 2D graphene oxide films integrated on silicon photonic chips", *Advanced Materials* Vol. 36 2403659 (2024). DOI: 10.1002/adma.202403659.
66. Di Jin, Jiayang Wu, Junkai Hu, Wenbo Liu1, Yuning Zhang, Yunyi Yang, Linnan Jia, Duan Huang, Baohua Jia, and David J. Moss, "Silicon photonic waveguide and microring resonator polarizers incorporating 2D graphene oxide films", *Applied Physics Letters* vol. 125, 000000 (2024); doi: 10.1063/5.0221793.
67. Yuning Zhang, Jiayang Wu, Linnan Jia, Di Jin, Baohua Jia, Xiaoyong Hu, David Moss, Qihuang Gong, "Advanced optical polarizers based on 2D materials", *npj Nanophotonics* Vol. 1, (2024). DOI: 10.1038/s44310-024-00028-3.
68. Yang Qu, Jiayang Wu, Yuning Zhang, Yunyi Yang, Linnan Jia, Houssein El Dirani, Sébastien Kerdiles, Corrado Sciancalepore, Pierre Demongodin, Christian Grillet, Christelle Monat, Baohua Jia, and David J. Moss, "Integrated optical parametric amplifiers in silicon nitride waveguides incorporated with 2D graphene oxide films", *Light: Advanced Manufacturing* **4** 39 (2023). https://doi.org/10.37188/lam.2023.039.
69. Di Jin, Wenbo Liu, Linnan Jia, Junkai Hu, Duan Huang, Jiayang Wu, Baohua Jia, and David J. Moss, "Thickness and Wavelength Dependent Nonlinear Optical Absorption in 2D Layered MXene Films", *Small Science* **4** 2400179 (2024). DOI:10.1002/smsc202400179;
70. Linnan Jia, Jiayang Wu, Yuning Zhang, Yang Qu, Baohua Jia, Zhigang Chen, and David J. Moss, "Fabrication Technologies for the On-Chip Integration of 2D Materials", Small: Methods Vol. 6, 2101435 (2022). DOI:10.1002/smtd.202101435.
71. Yuning Zhang, Jiayang Wu, Yang Qu, Linnan Jia, Baohua Jia, and David J. Moss, "Design and optimization of four-wave mixing in microring resonators integrated with 2D graphene oxide films", Journal of Lightwave Technology Vol. 39 (20) 6553-6562 (2021). DOI:10.1109/JLT.2021.3101292.
72. Yuning Zhang, Jiayang Wu, Yang Qu, Linnan Jia, Baohua Jia, and David J. Moss, "Optimizing the Kerr nonlinear optical performance of silicon waveguides integrated with 2D graphene oxide films", Journal of Lightwave Technology Vol. 39 (14) 4671-4683 (2021). DOI: 10.1109/JLT.2021.3069733.
73. Di Jin, Jiayang Wu, Junkai Hu, Wenbo Liu1, Yuning Zhang, Yunyi Yang, Linnan Jia, Duan Huang, Baohua Jia, and David J. Moss, "Silicon photonic waveguide and microring resonator polarizers incorporating 2D graphene oxide films", Applied Physics Letters Vol. 125, 053101 (2024). doi: 10.1063/5.0221793.
74. Yang Qu, Jiayang Wu, Yuning Zhang, Yao Liang, Baohua Jia, and David J. Moss, "Analysis of four-wave mixing in silicon nitride waveguides integrated with 2D layered graphene oxide films", Journal of Lightwave Technology Vol. 39 (9) 2902-2910 (2021). DOI: 10.1109/JLT.2021.3059721.

75. Jiayang Wu, Linnan Jia, Yuning Zhang, Yang Qu, Baohua Jia, and David J. Moss, "Graphene oxide: versatile films for flat optics to nonlinear photonic chips", Advanced Materials Vol. 33 (3) 2006415, pp.1-29 (2021). DOI:10.1002/adma.202006415.
76. Y. Qu, J. Wu, Y. Zhang, L. Jia, Y. Yang, X. Xu, S. T. Chu, B. E. Little, R. Morandotti, B. Jia, and D. J. Moss, "Graphene oxide for enhanced optical nonlinear performance in CMOS compatible integrated devices", Paper No. 11688-30, PW21O-OE109-36, 2D Photonic Materials and Devices IV, SPIE Photonics West, San Francisco CA March 6-11 (2021). doi.org/10.1117/12.2583978
77. Jiayang Wu, Yunyi Yang, Yang Qu, Xingyuan Xu, Yao Liang, Sai T. Chu, Brent E. Little, Roberto Morandotti, Baohua Jia, and David J. Moss, "Graphene oxide waveguide polarizers and polarization selective micro-ring resonators", Paper 11282-29, SPIE Photonics West, San Francisco, CA, 4 - 7 February (2020). doi: 10.1117/12.2544584
78. Yunyi Yang, Jiayang Wu, Xingyuan Xu, Sai T. Chu, Brent E. Little, Roberto Morandotti, Baohua Jia, and David J. Moss, "Enhanced four-wave mixing in graphene oxide coated waveguides", Applied Physics Letters Photonics vol. 3 120803 (2018). doi: 10.1063/1.5045509.
79. Linnan Jia, Yang Qu, Jiayang Wu, Yuning Zhang, Yunyi Yang, Baohua Jia, and David J. Moss, "Third-order optical nonlinearities of 2D materials at telecommunications wavelengths", Micromachines, **14**, 307 (2023). https://doi.org/10.3390/mi14020307.
80. Linnan Jia, Dandan Cui, Jiayang Wu, Haifeng Feng, Tieshan Yang, Yunyi Yang, Yi Du, Weichang Hao, Baohua Jia, David J. Moss, "BiOBr nanoflakes with strong nonlinear optical properties towards hybrid integrated photonic devices", Applied Physics Letters Photonics vol. 4 090802 vol. (2019). DOI: 10.1063/1.5116621
81. Linnan Jia, Jiayang Wu, Yunyi Yang, Yi Du, Baohua Jia, David J. Moss, "Large Third-Order Optical Kerr Nonlinearity in Nanometer-Thick PdSe2 2D Dichalcogenide Films: Implications for Nonlinear Photonic Devices", ACS Applied Nano Materials vol. 3 (7) 6876–6883 (2020). DOI:10.1021/acsanm.0c01239.
82. Moss, D. J., Morandotti, R., Gaeta, A. L. & Lipson, M. New CMOS compatible platforms based on silicon nitride and Hydex for nonlinear optics. Nat. Photonics Vol. 7, 597-607 (2013).
83. L. Razzari, et al., "CMOS-compatible integrated optical hyper-parametric oscillator," Nature Photonics, vol. 4, no. 1, pp. 41-45, 2010.
84. A. Pasquazi, et al., "Sub-picosecond phase-sensitive optical pulse characterization on a chip", Nature Photonics, vol. 5, no. 10, pp. 618-623 (2011).
85. M Ferrera et al., "On-Chip ultra-fast 1st and 2nd order CMOS compatible all-optical integration", Optics Express vol. 19 (23), 23153-23161 (2011).
86. Bao, C., et al., Direct soliton generation in microresonators, Opt. Lett, 42, 2519 (2017).

87. M.Ferrera et al., "CMOS compatible integrated all-optical RF spectrum analyzer", Optics Express, vol. 22, no. 18, 21488 - 21498 (2014).
88. M. Kues, et al., "Passively modelocked laser with an ultra-narrow spectral width", Nature Photonics, vol. 11, no. 3, pp. 159, 2017.
89. M. Ferrera, et al., "Low-power continuous-wave nonlinear optics in doped silica glass integrated waveguide structures," Nature Photonics, vol. 2, no. 12, pp. 737-740, 2008.
90. M.Ferrera et al."On-Chip ultra-fast 1st and 2nd order CMOS compatible all-optical integration", Opt. Express, vol. 19, (23)pp. 23153-23161 (2011).
91. D. Duchesne, M. Peccianti, M. R. E. Lamont, et al., "Supercontinuum generation in a high index doped silica glass spiral waveguide," Optics Express, vol. 18, no, 2, pp. 923-930, 2010.
92. H Bao, L Olivieri, M Rowley, ST Chu, BE Little, R Morandotti, DJ Moss, ... "Turing patterns in a fiber laser with a nested microresonator: Robust and controllable microcomb generation", Physical Review Research vol. 2 (2), 023395 (2020).
93. M. Ferrera, et al., "On-chip CMOS-compatible all-optical integrator", Nature Communications, vol. 1, Article 29, 2010.
94. A. Pasquazi, et al., "All-optical wavelength conversion in an integrated ring resonator," Optics Express, vol. 18, no. 4, pp. 3858-3863, 2010.
95. A.Pasquazi, Y. Park, J. Azana, et al., "Efficient wavelength conversion and net parametric gain via Four Wave Mixing in a high index doped silica waveguide," Optics Express, vol. 18, no. 8, pp. 7634-7641, 2010.
96. Peccianti, M. Ferrera, L. Razzari, et al., "Subpicosecond optical pulse compression via an integrated nonlinear chirper," Optics Express, vol. 18, no. 8, pp. 7625-7633, 2010.
97. M Ferrera, Y Park, L Razzari, BE Little, ST Chu, R Morandotti, DJ Moss, ... et al., "All-optical 1st and 2nd order integration on a chip", Optics Express vol. 19 (23), 23153-23161 (2011).
98. M. Ferrera et al., "Low Power CW Parametric Mixing in a Low Dispersion High Index Doped Silica Glass Micro-Ring Resonator with Q-factor > 1 Million", Optics Express, vol.17, no. 16, pp. 14098–14103 (2009).
99. M. Peccianti, et al., "Demonstration of an ultrafast nonlinear microcavity modelocked laser", Nature Communications, vol. 3, pp. 765, 2012.
100. A.Pasquazi, et al., "Self-locked optical parametric oscillation in a CMOS compatible microring resonator: a route to robust optical frequency comb generation on a chip," Optics Express, vol. 21, no. 11, pp. 13333-13341, 2013.
101. A.Pasquazi, et al., "Stable, dual mode, high repetition rate mode-locked laser based on a microring resonator," Optics Express, vol. 20, no. 24, pp. 27355-27362, 2012.
102. Pasquazi, A. et al. Micro-combs: a novel generation of optical sources. Physics Reports 729, 1-81 (2018).
103. Yang Sun, Jiayang Wu, Mengxi Tan, Xingyuan Xu, Yang Li, Roberto Morandotti, Arnan Mitchell, and David J. Moss, "Applications of optical micro-combs", Advances in Optics and Photonics **15** (1) 86-175 (2023). https://doi.org/10.1364/AOP.470264.

104. H. Bao, et al., Laser cavity-soliton microcombs, Nature Photonics, vol. 13, no. 6, pp. 384-389, Jun. 2019.
105. Antonio Cutrona, Maxwell Rowley, Debayan Das, Luana Olivieri, Luke Peters, Sai T. Chu, Brent L. Little, Roberto Morandotti, David J. Moss, Juan Sebastian Totero Gongora, Marco Peccianti, Alessia Pasquazi, "High Conversion Efficiency in Laser Cavity-Soliton Microcombs", Optics Express Vol. 30, Issue 22, pp. 39816-39825 (2022). https://doi.org/10.1364/OE.470376.
106. M.Rowley, P.Hanzard, A.Cutrona, H.Bao, S.Chu, B.Little, R.Morandotti, D. J. Moss, G. Oppo, J. Gongora, M. Peccianti and A. Pasquazi, "Self-emergence of robust solitons in a micro-cavity", Nature vol. 608 (7922) 303–309 (2022).
107. A. Cutrona, M. Rowley, A. Bendahmane, V. Cecconi, L. Peters, L. Olivieri, B. E. Little, S. T. Chu, S. Stivala, R. Morandotti, D. J. Moss, J. S. Totero-Gongora, M. Peccianti, A. Pasquazi, "Nonlocal bonding of a soliton and a blue-detuned state in a microcomb laser", *Nature Communications Physics* **6** Article 259 (2023). https://doi.org/10.1038/s42005-023-01372-0.
108. Aadhi A. Rahim, Imtiaz Alamgir, Luigi Di Lauro, Bennet Fischer, Nicolas Perron, Pavel Dmitriev, Celine Mazoukh, Piotr Roztocki, Cristina Rimoldi, Mario Chemnitz, Armaghan Eshaghi, Evgeny A. Viktorov, Anton V. Kovalev, Brent E. Little, Sai T. Chu, David J. Moss, and Roberto Morandotti, "Mode-locked laser with multiple timescales in a microresonator-based nested cavity", *APL Photonics* **9** 031302 (2024). DOI:10.1063/5.0174697.
109. A. Cutrona, M. Rowley, A. Bendahmane, V. Cecconi, L. Peters, L. Olivieri, B. E. Little, S. T. Chu, S. Stivala, R. Morandotti, D. J. Moss, J. S. Totero-Gongora, M. Peccianti, A. Pasquazi, "Stability Properties of Laser Cavity-Solitons for Metrological Applications", Applied Physics Letters vol. 122 (12) 121104 (2023); doi: 10.1063/5.0134147.
110. X. Xu, J. Wu, M. Shoeiby, T. G. Nguyen, S. T. Chu, B. E. Little, R. Morandotti, A. Mitchell, and D. J. Moss, "Reconfigurable broadband microwave photonic intensity differentiator based on an integrated optical frequency comb source," APL Photonics, vol. 2, no. 9, 096104, Sep. 2017.
111. Xu, X., et al., Photonic microwave true time delays for phased array antennas using a 49 GHz FSR integrated micro-comb source, Photonics Research, vol. 6, B30-B36 (2018).
112. X. Xu, M. Tan, J. Wu, R. Morandotti, A. Mitchell, and D. J. Moss, "Microcomb-based photonic RF signal processing", IEEE Photonics Technology Letters, vol. 31 no. 23 1854-1857, 2019.
113. Xu, et al., "Advanced adaptive photonic RF filters with 80 taps based on an integrated optical micro-comb source," Journal of Lightwave Technology, vol. 37, no. 4, pp. 1288-1295 (2019).
114. X. Xu, et al., "Photonic RF and microwave integrator with soliton crystal microcombs", IEEE Transactions on Circuits and Systems II: Express Briefs, vol. 67, no. 12, pp. 3582-3586, 2020. DOI:10.1109/TCSII.2020.2995682.
115. X. Xu, et al., "High performance RF filters via bandwidth scaling with Kerr micro-combs," APL Photonics, vol. 4 (2) 026102. 2019.

116. M. Tan, et al., "Microwave and RF photonic fractional Hilbert transformer based on a 50 GHz Kerr micro-comb", Journal of Lightwave Technology, vol. 37, no. 24, pp. 6097 – 6104, 2019.
117. M. Tan, et al., "RF and microwave fractional differentiator based on photonics", IEEE Transactions on Circuits and Systems: Express Briefs, vol. 67, no.11, pp. 2767-2771, 2020. DOI:10.1109/TCSII.2020.2965158.
118. M. Tan, et al., "Photonic RF arbitrary waveform generator based on a soliton crystal micro-comb source", Journal of Lightwave Technology, vol. 38, no. 22, pp. 6221-6226 (2020). DOI: 10.1109/JLT.2020.3009655.
119. M. Tan, X. Xu, J. Wu, R. Morandotti, A. Mitchell, and D. J. Moss, "RF and microwave high bandwidth signal processing based on Kerr Micro-combs", Advances in Physics X, VOL. 6, NO. 1, 1838946 (2021). DOI:10.1080/23746149.2020.1838946.
120. X. Xu, et al., "Advanced RF and microwave functions based on an integrated optical frequency comb source," Opt. Express, vol. 26 (3) 2569 (2018).
121. M. Tan, X. Xu, J. Wu, B. Corcoran, A. Boes, T. G. Nguyen, S. T. Chu, B. E. Little, R. Morandotti, A. Lowery, A. Mitchell, and D. J. Moss, ""Highly Versatile Broadband RF Photonic Fractional Hilbert Transformer Based on a Kerr Soliton Crystal Microcomb", Journal of Lightwave Technology vol. 39 (24) 7581-7587 (2021).
122. Wu, J. et al. RF Photonics: An Optical Microcombs' Perspective. IEEE Journal of Selected Topics in Quantum Electronics Vol. 24, 6101020, 1-20 (2018).
123. T. G. Nguyen et al., "Integrated frequency comb source-based Hilbert transformer for wideband microwave photonic phase analysis," Opt. Express, vol. 23, no. 17, pp. 22087-22097, Aug. 2015.
124. X. Xu, et al., "Broadband RF channelizer based on an integrated optical frequency Kerr comb source," Journal of Lightwave Technology, vol. 36, no. 19, pp. 4519-4526, 2018.
125. X. Xu, et al., "Continuously tunable orthogonally polarized RF optical single sideband generator based on micro-ring resonators," Journal of Optics, vol. 20, no. 11, 115701. 2018.
126. X. Xu, et al., "Orthogonally polarized RF optical single sideband generation and dual-channel equalization based on an integrated microring resonator," Journal of Lightwave Technology, vol. 36, no. 20, pp. 4808-4818. 2018.
127. X. Xu, et al., "Photonic RF phase-encoded signal generation with a microcomb source", J. Lightwave Technology, vol. 38, no. 7, 1722-1727, 2020.
128. X. Xu, et al., Broadband microwave frequency conversion based on an integrated optical micro-comb source", Journal of Lightwave Technology, vol. 38 no. 2, pp. 332-338, 2020.
129. M. Tan, et al., "Photonic RF and microwave filters based on 49GHz and 200GHz Kerr microcombs", Optics Comm. vol. 465,125563, Feb. 22. 2020.
130. X. Xu, et al., "Broadband photonic RF channelizer with 90 channels based on a soliton crystal microcomb", Journal of Lightwave Technology, Vol. 38, no. 18, pp. 5116 – 5121 (2020). doi: 10.1109/JLT.2020.2997699.
131. M. Tan et al, "Orthogonally polarized Photonic Radio Frequency single sideband generation with integrated micro-ring resonators", IOP Journal of

Semiconductors, Vol. 42 (4), 041305 (2021). DOI: 10.1088/1674-4926/42/4/041305.

132. Mengxi Tan, X. Xu, J. Wu, T. G. Nguyen, S. T. Chu, B. E. Little, R. Morandotti, A. Mitchell, and David J. Moss, "Photonic Radio Frequency Channelizers based on Kerr Optical Micro-combs", IOP Journal of Semiconductors Vol. 42 (4), 041302 (2021). DOI:10.1088/1674-4926/42/4/041302.

133. B. Corcoran, et al., "Ultra-dense optical data transmission over standard fiber with a single chip source", Nature Communications, vol. 11, Article:2568, 2020.

134. X. Xu et al, "Photonic perceptron based on a Kerr microcomb for scalable high speed optical neural networks", Laser and Photonics Reviews, vol. 14, no. 8, 2000070 (2020). DOI: 10.1002/lpor.202000070.

135. Xingyuan Xu, Weiwei Han, Mengxi Tan, Yang Sun, Yang Li, Jiayang Wu, Roberto Morandotti, Arnan Mitchell, Kun Xu, and David J. Moss, "Neuromorphic computing based on wavelength-division multiplexing", ***IEEE Journal of Selected Topics in Quantum Electronics* 29** (2) 7400112 (2023). DOI:10.1109/JSTQE.2022.3203159.

136. Yunping Bai, Xingyuan Xu,1, Mengxi Tan, Yang Sun, Yang Li, Jiayang Wu, Roberto Morandotti, Arnan Mitchell, Kun Xu, and David J. Moss, "Photonic multiplexing techniques for neuromorphic computing", Nanophotonics vol. 12 (5): 795–817 (2023). DOI:10.1515/nanoph-2022-0485.

137. Chawaphon Prayoonyong, Andreas Boes, Xingyuan Xu, Mengxi Tan, Sai T. Chu, Brent E. Little, Roberto Morandotti, Arnan Mitchell, David J. Moss, and Bill Corcoran, "Frequency comb distillation for optical superchannel transmission", Journal of Lightwave Technology vol. 39 (23) 7383-7392 (2021). DOI: 10.1109/JLT.2021.3116614.

138. Mengxi Tan, Xingyuan Xu, Jiayang Wu, Bill Corcoran, Andreas Boes, Thach G. Nguyen, Sai T. Chu, Brent E. Little, Roberto Morandotti, Arnan Mitchell, and David J. Moss, "Integral order photonic RF signal processors based on a soliton crystal micro-comb source", IOP Journal of Optics vol. 23 (11) 125701 (2021). https://doi.org/10.1088/2040-8986/ac2eab

139. Yang Sun, Jiayang Wu, Yang Li, Xingyuan Xu, Guanghui Ren, Mengxi Tan, Sai Tak Chu, Brent E. Little, Roberto Morandotti, Arnan Mitchell, and David J. Moss, "Optimizing the performance of microcomb based microwave photonic transversal signal processors", Journal of Lightwave Technology vol. 41 (23) pp 7223-7237 (2023). DOI: 10.1109/JLT.2023.3314526.

140. Mengxi Tan, Xingyuan Xu, Andreas Boes, Bill Corcoran, Thach G. Nguyen, Sai T. Chu, Brent E. Little, Roberto Morandotti, Jiayang Wu, Arnan Mitchell, and David J. Moss, "Photonic signal processor for real-time video image processing based on a Kerr microcomb", Communications Engineering vol. 2 94 (2023). DOI:10.1038/s44172-023-00135-7

141. Yang Sun, Jiayang Wu, Yang Li, Mengxi Tan, Xingyuan Xu, Sai Chu, Brent Little, Roberto Morandotti, Arnan Mitchell, and David J. Moss, "Quantifying the Accuracy of Microcomb-based Photonic RF Transversal Signal Processors", IEEE Journal of Selected Topics in Quantum Electronics vol. 29 no. 6, pp. 1-17, Art no. 7500317 (2023). 10.1109/JSTQE.2023.3266276.

142. Yang Li, Yang Sun, Jiayang Wu, Guanghui Ren, Bill Corcoran, Xingyuan Xu, Sai

T. Chu, Brent. E. Little, Roberto Morandotti, Arnan Mitchell, and David J. Moss, "Processing accuracy of microcomb-based microwave photonic signal processors for different input signal waveforms", ***MDPI Photonics*** **10**, 10111283 (2023). DOI:10.3390/photonics10111283

143. Caitlin E. Murray, Mengxi Tan, Chawaphon Prayoonyong, Sai T. Chu, Brent E. Little, Roberto Morandotti, Arnan Mitchell, David J. Moss and Bill Corcoran, "Investigating the thermal robustness of soliton crystal microcombs", ***Optics Express*** **31**(23), 37749-37762 (2023).

144. Yang Sun, Jiayang Wu, Yang Li, and David J. Moss, "Comparison of microcomb-based RF photonic transversal signal processors implemented with discrete components versus integrated chips", ***MDPI Micromachines*** **14**, 1794 (2023). https://doi.org/10.3390/mi14091794

145. C. Mazoukh, L. Di Lauro, I. Alamgir1 B. Fischer, A. Aadhi, A. Eshaghi, B. E. Little, S. T. Chu, D. J. Moss, and R. Morandotti, "Genetic algorithm-enhanced microcomb state generation", ***Nature Communications Physics*** Vol. 7, Article: 81 (2024). DOI: 10.1038/s42005-024-01558-0.

146. Mengxi Tan, David J. Moss, "The laser trick that could put an ultraprecise optical clock on a chip", ***Nature*** **624**, (7991) 256-257 (2023). *doi.org/10.1038/d41586-023-03782-0.*

147. Andrew Cooper, Luana Olivieri, Antonio Cutrona, Debayan Das, Luke Peters, Sai Tak Chu, Brent Little, Roberto Morandotti, David J Moss, Marco Peccianti, and Alessia Pasquazi, "Parametric interaction of laser cavity-solitons with an external CW pump", ***Optics Express*** **32** (12), 21783-21794 (2024).

148. Weiwei Han, Zhihui Liu, Yifu Xu, Mengxi Tan, Chaoran Huang, Jiayang Wu, Kun Xu, David J. Moss, and Xingyuan Xu, "Photonic RF Channelization Based on Microcombs", ***IEEE Journal of Selected Topics in Quantum Electronics*** **30** (5) 7600417 (2024). DOI:10.1109/JSTQE.2024.3398419.

149. Yang Li, Yang Sun, Jiayang Wu, Guanghui Ren, Xingyuan Xu, Mengxi Tan, Sai Chu, Brent Little, Roberto Morandotti, Arnan Mitchell, and David Moss, "Feedback control in micro-comb-based microwave photonic transversal filter systems", ***IEEE Journal of Selected Topics in Quantum Electronics*** Vol. **30** (5) 2900117 (2024). DOI: 10.1109/JSTQE.2024.3377249.

150. Weiwei Han, Zhihui Liu, Yifu Xu, Mengxi Tan, Yuhua Li, Xiaotian Zhu, Yanni Ou, Feifei Yin, Roberto Morandotti, Brent E. Little, Sai Tak Chu, Xingyuan Xu, David J. Moss, and Kun Xu, "Dual-polarization RF Channelizer Based on Microcombs", ***Optics Express*** **32**, No. 7, 11281-11295 (2024). DOI: 10.1364/OE.519235.

151. Mengxi Tan, Xingyuan Xu, Andreas Boes, Bill Corcoran, Thach G. Nguyen, Sai T. Chu, Brent E. Little, Roberto Morandotti, Jiayang Wu, Arnan Mitchell, and David J. Moss, "Photonic signal processor for real-time video image processing based on a Kerr microcomb", ***Nature Communications Engineering*** **2** 94 (2023). DOI:10.1038/s44172-023-00135-7.

152. H. Yu, S. Sciara, M. Chemnitz, N. Montaut, B. Fischer, R. Helsten, B. Crockett, B. Wetzel, T. A. Göbel, R. Krämer, B. E. Little, S. T. Chu, D. J. Moss, S. Nolte, W.J. Munro, Z. Wang, J. Azaña, R. Morandotti, "Quantum key distribution implemented with d-level time-bin entangled photons", Nature Communications Vol. 16, 171 (2025). DOI:10.1038/s41467-024-55345-0.
153. Weiwei Han, Zhihui Liu, Yifu Xu, Mengxi Tan, Yuhua Li, Xiaotian Zhu, Yanni Ou, Feifei Yin, Roberto Morandotti, Brent E. Little, Sai Tak Chu, David J. Moss, Xingyuan Xu, and Kun Xu, "TOPS-speed complex-valued convolutional accelerator for feature extraction and inference", Nature Communications Vol. 16, 292 (2025). DOI: 10.1038/s41467-024-55321-8.
154. Zhihui Liu, Haoran Zhang, Yuhang Song, Xiaotian Zhu, Yunping Bai, Mengxi Tan, Bill Corcoran, Caitlin Murphy, Sai T. Chu, David J. Moss, Xingyuan Xu, and Kun Xu, "Advances in Soliton Crystal Microcombs", Photonics Vol. 11, 1164 (2024). https://doi.org/10.3390/photonics11121164.
155. Bill Corcoran, Arnan Mitchell, Roberto Morandotti, Leif K. Oxenlowe, and David J. Moss, "Microcombs for Optical Communications", **Nature Photonics 19** (2025).
156. Yang Li, Yang Sun, Jiayang Wu, Guanghui Ren, Roberto Morandotti, Xingyuan Xu, Mengxi Tan, Arnan Mitchell, and David J. Moss, "Performance analysis of microwave photonic spectral filters based on optical microcombs", **Advanced Physics Research 3** (9) (2024). DOI:10.1002/apxr.202400084.
157. Aadhi A. Rahim, Imtiaz Alamgir, Luigi Di Lauro, Bennet Fischer, Nicolas Perron, Pavel Dmitriev, Celine Mazoukh, Piotr Roztocki, Cristina Rimoldi, Mario Chemnitz, Armaghan Eshaghi, Evgeny A. Viktorov, Anton V. Kovalev, Brent E. Little, Sai T. Chu, David J. Moss, and Roberto Morandotti, "Mode-locked laser with multiple timescales in a microresonator-based nested cavity", **APL Photonics 9**  031302 (2024); DOI:10.1063/5.0174697.
158. Yonghang Sun, James Salamy, Caitlin E. Murry, Brent E. Little, Sai T. Chu, Roberto Morandotti, Arnan Mitchell, David J. Moss, Bill Corcoran, "Enhancing laser temperature stability by passive self-injection locking to a micro-ring resonator", **Optics Express 32** (13) 23841-23855 (2024) https://doi.org/10.1364/OE.515269.
159. J. Capmany and D. Novak, "Microwave photonics combines two worlds," Nat. Photonics, vol. 1, no. 6, pp. 319-330, 2007/06/01 2007, doi: 10.1038/nphoton.2007.89.
160. J. Yao, "Microwave Photonics," Journal of Lightwave Technology, vol. 27, no. 3, pp. 314-335, 2009/02/01 2009, doi: 10.1109/JLT.2008.2009551.
161. D. Marpaung, J. Yao, and J. Capmany, "Integrated microwave photonics," Nat. Photonics, vol. 13, no. 2, pp. 80-90, 2019/02/01 2019, doi: 10.1038/s41566-018-0310-5.
162. J. Capmany, B. Ortega, and D. Pastor, "A Tutorial on Microwave Photonic Filters," Journal of Lightwave Technology, vol. 24, no. 1, p. 201, 2006/01/01 2006, doi: 10.1109/JLT.2005.860478.
163. Y. Liu, A. Choudhary, D. Marpaung, and B. J. Eggleton, "Integrated microwave photonic filters," Adv. Opt. Photon., vol. 12, no. 2, pp. 485-555, 2020/06/30 2020, doi: 10.1364/AOP.378686.
164. Y. Long and J. Wang, "Ultra-high peak rejection notch microwave photonic filter using a single silicon microring resonator," Opt. Express, vol. 23, no. 14, pp. 17739-17750, 2015/07/13 2015, doi: 10.1364/OE.23.017739.

165. A. Choudhary et al., "Advanced Integrated Microwave Signal Processing With Giant On-Chip Brillouin Gain," Journal of Lightwave Technology, vol. 35, no. 4, pp. 846-854, 2017, doi: 10.1109/JLT.2016.2613558.
166. Z. Zhu, H. Chi, T. Jin, S. Zheng, X. Jin, and X. Zhang, "All-positive-coefficient microwave photonic filter with rectangular response," Opt. Lett., vol. 42, no. 15, pp. 3012-3015, 2017/08/01 2017, doi: 10.1364/OL.42.003012.
167. Hamed Arianfard, Jiayang Wu, Saulius Juodkazis, and David J. Moss, "Optical analogs of Rabi splitting in integrated waveguide-coupled resonators", Advanced Physics Research  2 (2023). DOI: 10.1002/apxr.202200123.
168. Hamed Arianfard, Jiayang Wu, Saulius Juodkazis, and David J. Moss, "Spectral shaping based on optical waveguides with advanced Sagnac loop reflectors", Paper No. PW22O-OE201-20, SPIE-Opto, Integrated Optics: Devices, Materials, and Technologies XXVI, SPIE Photonics West, San Francisco CA January 22 - 27 (2022).  doi: 10.1117/12.2607902
169. Hamed Arianfard, Jiayang Wu, Saulius Juodkazis, David J. Moss, "Spectral Shaping Based on Integrated Coupled Sagnac Loop Reflectors Formed by a Self-Coupled Wire Waveguide", IEEE Photonics Technology Letters vol. 33 (13) 680-683 (2021). DOI:10.1109/LPT.2021.3088089.
170. Hamed Arianfard, Jiayang Wu, Saulius Juodkazis and David J. Moss, "Three Waveguide Coupled Sagnac Loop Reflectors for Advanced Spectral Engineering", Journal of Lightwave Technology vol. 39 (11) 3478-3487 (2021). DOI: 10.1109/JLT.2021.3066256.
171. Hamed Arianfard, Jiayang Wu, Saulius Juodkazis and David J. Moss, "Advanced Multi-Functional Integrated Photonic Filters based on Coupled Sagnac Loop Reflectors", Journal of Lightwave Technology vol. 39 Issue: 5, pp.1400-1408 (2021). DOI:10.1109/JLT.2020.3037559.
172. Hamed Arianfard, Jiayang Wu, Saulius Juodkazis and David J. Moss, "Advanced multi-functional integrated photonic filters based on coupled Sagnac loop reflectors", Paper 11691-4, PW21O-OE203-44, Silicon Photonics XVI, SPIE Photonics West, San Francisco CA March 6-11 (2021). doi.org/10.1117/12.2584020
173. Jiayang Wu, Tania Moein, Xingyuan Xu, and David J. Moss, "Advanced photonic filters via cascaded Sagnac loop reflector resonators in silicon-on-insulator integrated nanowires", Applied Physics Letters Photonics vol. 3 046102 (2018). DOI:/10.1063/1.5025833
174. Jiayang Wu, Tania Moein, Xingyuan Xu, Guanghui Ren, Arnan Mitchell, and David J. Moss, "Micro-ring resonator quality factor enhancement via an integrated Fabry-Perot cavity", Applied Physics Letters Photonics vol. 2 056103 (2017). doi: 10.1063/1.4981392.